\documentclass[reprint,
floatfix,
 amsmath,amssymb,
 aps,
 prl,superscriptaddress,longbibliography]{revtex4-1}

\usepackage{graphicx}%
\usepackage{dcolumn}%
\usepackage{bm}%
\usepackage[pdftex]{hyperref}%
\usepackage{bbold}
\usepackage{dsfont}
\usepackage{amsthm} %
\usepackage{braket}
\usepackage{natbib}
\usepackage{float}
\usepackage{multirow}
\usepackage[table]{xcolor}
\usepackage{tikz}
\usepackage{comment}

\def\lb{\left(}
\def\rb{\right)}
\newcommand{\cO}{\mathcal{O}}

\newcommand{\C}{\mathds{C}}

\newcommand{\ketbra}[2]{|#1\rangle\langle #2|}
\newcommand{\scalar}[2]{\langle #1, #2\rangle}

\newcommand{\bbC}{\mathbb{C}}
\newcommand{\calA}{\mathcal{A}}
\newcommand{\calB}{\mathcal{B}}

\newcommand{\frakd}{\mathfrak{d}}
\newcommand{\textsum}{{\textstyle \sum}}
\newcommand{\textbigotimes}{{\textstyle \bigotimes}}
\newcommand{\vvirg}{, \dots ,}
\newcommand{\eps}{\epsilon}
\DeclareMathOperator{\bond}{bond}
\DeclareMathOperator{\bbond}{\underline{bond}}
\DeclareMathOperator{\TNS}{TNS}
\DeclareMathOperator{\bTNS}{bTNS}
\DeclareMathOperator{\cTNS}{\bar{TNS}}
\DeclareMathOperator{\ghz}{ghz}
\renewcommand{\tilde}[1]{\widetilde{#1}}
\renewcommand{\bar}[1]{\overline{#1}}

\newtheorem*{theorem*}{Theorem}

\newtheorem*{lem*}{Lemma}

 \newcommand{\tns}{\mathrm{tns}}
 \newcommand{\btns}{\mathrm{btns}}
\begin{document}

\title{Optimization at the boundary of the tensor network variety}%

 \author{Matthias Christandl}
\email{christandl@math.ku.dk}
 \affiliation{QMATH, Department of Mathematical Sciences, University of Copenhagen, Universitetsparken 5, 2100 Copenhagen, Denmark}

 \author{Fulvio Gesmundo}
 \email{fulvio.gesmundo@mis.mpg.de}
 \affiliation{QMATH, Department of Mathematical Sciences, University of Copenhagen, Universitetsparken 5, 2100 Copenhagen, Denmark}
 \affiliation{Max Planck Institute for Mathematics in the Science, Inselstrasse 22, 04103 Leipzig, Germany}

\author{Daniel Stilck Fran\c{c}a}
 \email{dsfranca@math.ku.dk}
 \affiliation{QMATH, Department of Mathematical Sciences, University of Copenhagen, Universitetsparken 5, 2100 Copenhagen, Denmark}

 \author{Albert H. Werner}
\email{werner@math.ku.dk}
\affiliation{QMATH, Department of Mathematical Sciences, University of Copenhagen, Universitetsparken 5, 2100 Copenhagen, Denmark}
\affiliation{NBIA, Niels Bohr Institute, University of Copenhagen, Blegdamsvej 17, 2100 Copenhagen, Denmark}

\date{\today}

\begin{abstract}
Tensor network states form a variational ansatz class widely used, both analytically and numerically, in the study of quantum many-body systems. It is known that if the underlying graph contains a cycle, e.g. as in projected entangled pair states (PEPS), then the set of tensor network states of given bond dimension is not closed. Its closure is the tensor network variety. Recent work has shown that states on the boundary of this variety can yield more efficient representations for states of physical interest, but it remained unclear how to systematically find and optimize over such representations. We address this issue by defining an ansatz class of states that includes states at the boundary of the tensor network variety of given bond dimension. We show how to optimize over this class in order to find ground states of local Hamiltonians by only slightly modifying standard algorithms and code for tensor networks. We apply this method to different models and observe favorable energies and runtimes when compared with standard tensor network methods. 
\end{abstract}

\maketitle

 Tensor network states are quantum states obtained by contracting tensors, placed on vertices of a graph, according to the edges of the graph that identify indices of the tensors. They are featured in successful approaches to the study of quantum and classical many-body systems, and in particular, they provide an efficient ansatz class for quantum many-body states satisfying an area law~\cite{White_1992,Fannes_1992,mapsgarcia,Shi_2006,Vidal_2008,Vidal_2007,Schuch_2010,verstraete2004renormalization,Hauschild_2018,Orus_practical,Landau_2015,Arad_2017,Schollw_ck_2011}. For a fixed graph, the expressive power of the ansatz class is determined by an integer parameter $D$, called \emph{bond dimension}. It is known that when the underlying graph contains cycles, the set of tensor network states of bond dimension at most $D$ is not closed in the standard Euclidean topology~\cite{Landsberg_geometry}; its closure is an algebraic variety called the \emph{tensor network variety}. In~\cite{christandl2018tensor}, the authors show that states of physical interest may belong to the boundary of the variety, i.e., their bond dimension is strictly higher than $D$ but they can be approximated arbitrarily well by states of bond dimension $D$. Moreover, they showed how to exploit these approximate representations to obtain more efficient representations of the target state. However, it remained unclear how to make a variational method out of such states.

In this article, we define an ansatz class which includes states on the boundary. This class can describe certain states that arise as ground states of local Hamiltonians more efficiently than the standard tensor network ansatz class. Although by definition states on the boundary can be approximated using standard methods, we argue that this gives rise to ill-conditioned tensors. This leads to a slower convergence in variational methods and requires a precision in the computations that scales with system size. Thus, although tensor networks are often only an effective approximation to many-body states and variational methods converge exponentially fast to a good approximation to the ground state interior, the convergence will be slow for boundary states. Such slow convergence has already been observed for frustrated systems on the Kagome lattice~\cite{xie_tensor_2014}, suggesting that the underlying states ground states lie on the boundary. In contrast, our methods remain stable and our examples show that they still converge fast to the target state while the traditional tensor network ansatz does not.

Standard numerical methods for tensor networks can be applied to this extended setting, with only slight modification to algorithms~\cite{Orus_practical,Schollw_ck_2011} and code: in particular, we show how to use the ansatz class to find better approximate representations of ground states compared to a standard tensor network ansatz. We demonstrate our methods numerically in different directions. Our ansatz class indeed achieves smaller energies with less runtime for Hamiltonians for which there is a provable separation for the bond dimension required to represent the ground state. In addition, the methods achieve smaller energies than an MPS ansatz with the same number of parameters for the Heisenberg chain on a ring.

Thus, our findings indicate that this ansatz class can be used to obtain better numerical results for models of physical interest, besides further advancing our understanding of the geometry of tensor network states. 

The paper is organized as follows. We will first discuss the structure of the states on the boundary. In the next step we define an ansatz class which includes those states. We then discuss how to perform computations with states from this ansatz class before discussing the efficiency of their representation. The final part of the paper discusses the variation over this class and numerical examples comparing its performance to standard tensor network methods. A concluding discussion is provided at the end. 

\subparagraph{The structure of states on the boundary.} Let $G=(V,E)$ be a simple graph with a set of vertices $V$ and a set of edges $E$; let $L = |V|$ and let $D$ be a positive integer. For every edge  $y \in E$, let $\ket{\Omega_D^{(y)}}=\sum_{\alpha=1}^{D}\ket{\alpha,\alpha} \in \bbC^D \otimes \bbC^D$ be the unnormalized maximally entangled state of dimension $D$. Define $| \Omega ^{(G)}_D \rangle = \bigotimes_{e \in E} \ket{\Omega_D^{(e)}}$ and regard it as a vector in $\bigotimes_{v \in V} (\bbC^D)^{\otimes k_v}$, where $k_v$ is the degree of the vertex $v \in V$. Pictorially, this is the result of placing each $\ket{\Omega^{(e)}_D}$ on the corresponding edge of the graph and regarding the resulting tensor product $\ket{\Omega_D^{(G)}}$ as an unnormalized state on $L$ sites, corresponding to the vertices. 

For every $v \in V$, let $\mathcal{A}^v: \lb \C^{D}\rb^{\otimes k_v}\to \C^d$ be a linear map. Explicitly, write $\calA^v$ as a tensor 
\begin{align*}
 \mathcal{A}^v=\sum\limits_{i=1}^d\sum_{\alpha \in D ^{\times k_v}} A_{i,\alpha} \ketbra{i}{\alpha},
\end{align*}
where $D^{\times k_v} = \{ (\alpha_1 \vvirg \alpha_{k_v}) : 1 \leq \alpha_j \leq D\}$.

Given a family of linear maps $\calA = (\calA^v: v \in V)$, define $\tns^G(\calA) = \lb \textbigotimes_{v\in V}\mathcal{A}^v\rb \ket {\Omega_D^{(G)}}$. If $\ket{\psi} = \tns^G(\calA)$ for some choice of linear maps $\calA$, we say that $(\calA^v: v \in V)$ is a tensor network state representation for $\ket{\psi}$ with respect to $G$.

Let 
\[
\TNS^G_{D,d} = \bigl\{ \ket{\psi}= \tns^G ( \calA) : \calA = (\calA^v : v \in V) \bigr\}.
\]
For $\ket{\psi} \in (\bbC^d)^{\otimes L}$, the \emph{bond dimension} of $\ket{\psi}$ (with respect to the graph $G$) is defined as
\[
 \bond _{G}(\ket{\psi}) = \min \{ D : \ket{\psi} \in \TNS^G_{D,d}\}.
\]
We refer, e.g., to~\cite{verstraete2004renormalization} for details on this construction. It is straightforward to generalize this construction with different bond dimension $D$ at each edge, but we only consider the uniform case here for simplicity.
If $G$ is a ring, a tensor network representation is called a matrix product state representation (MPS)~\cite{mapsgarcia} with periodic boundary conditions; if $G$ is a lattice, it is called a projected entangled pair state representation (PEPS)~\cite{verstraete2004renormalization}. In addition, if the graph $G$ is regular and we have the same bond dimension at each edge, one can restrict to representations in which we apply the same map at each vertex. We call such tensor network representations \emph{translation invariant}.

If the graph $G$ contains cycles, then $\TNS_{D,d}^G$ is not closed, unless it coincides with the full $(\bbC^d)^{\otimes L}$~\cite{Landsberg_geometry}. Write $\cTNS_{D,d}^G$ for the closure of $\TNS_{D,d}^G$; the set $\cTNS_{D,d}^G$ is an algebraic variety and the (closure of the) difference $\cTNS_{D,d}^G \setminus \TNS_{D,d}^G$ is its algebraic \emph{boundary}. Following \cite{christandl2018tensor}, define the \emph{border bond dimension} of a state $\ket{\psi} \in (\bbC^d)^{\otimes L}$ as 
\[
\bbond_G(\ket{\psi})  = \min\{ D : \ket{\psi} \in \cTNS_{D,d}^G\}.
\]
Clearly $\bbond_G(\ket{\psi}) \leq \bond_G(\ket{\psi})$ and $\bbond_G(-)$ is a lower semicontinuous function. 

\begin{figure}[ht!]
\centering
\includegraphics[width=0.75\columnwidth]{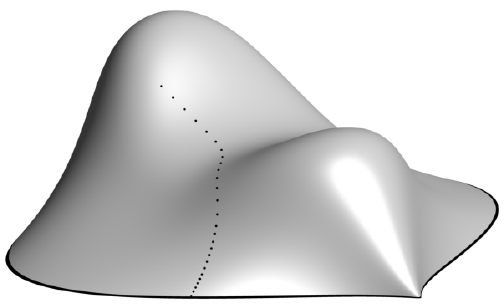}
\caption{Schematic representation of a curve converging to the boundary of the variety. The gray surface represents $\TNS^G_{D,d}$, the dotted black curve represents a sequence $\epsilon^{-a}\ket{\phi(\epsilon)}$ converging to a state on the boundary, the thick black curve.}
\label{fig:exampledeg}
\end{figure}

Following \cite[Lemma 3.1.6.2]{Landsberg_2017}, it is easy to show that a state at the boundary can be approximated along a rational curve, see Figure \ref{fig:exampledeg}: in other words, if $\ket{\psi}$ lies on the boundary, then there exists a family of local maps $(\calA^v(\eps): v \in V)$ whose entries are polynomials of degree at most $\frakd$ in $\eps$, and an integer $a \geq 1$, such that 
\begin{align*}
     \ket{\psi(\epsilon)} = \tns^G(\calA(\eps)) =\epsilon^{a}\ket{\psi}+\textsum_{j=1}^{e}\epsilon^{a+j}\ket{\widetilde{\psi}_j};
 \end{align*}
note $\ket{\psi} = \lim_{\epsilon\to 0} \eps^{-a} \ket{\psi(\eps)}$. Following~\cite{christandl2018tensor}, we say that the state $\ket{\phi}$ is a \emph{degeneration} of $\ket{\Omega_D^{(G)}}$, and that it admits a border bond dimension $D$ representation; the integers $a$ and $e$ are called \emph{approximation degree} and \emph{error degree} respectively \cite{Christandl2017_TR,ChrGesJen:BorderRankNonMult}. 

Note that although $\eps^{-a}\ket{\psi(\eps)}$ converges, in general there is no sequence of linear maps $(\calA^v : v \in V)$ such that $\bigotimes_v \calA^v = \lim \eps^{-a} \bigotimes \calA^v(\eps)$; in this case, we say that the limit only exists \emph{non-locally} and we observe that the terms $\bigotimes \calA^v(\eps)$ of order lower than $a$ in $\eps$ vanish.

For a fixed $a$ and $\frakd$, we will define an ansatz class that allows one to optimize over boundary states admitting a border bond representation with approximation degree $a$ and local maps of degree at most $\frakd$ in $\eps$. We focus on the regime in which $a$ is of constant or linear order in the system size and $\frakd$ is constant. To the best of our knowledge, this encompasses all known examples. Moreover, one can always assume $\frakd \leq a$, as higher order terms in the local maps do not contribute to the limit.

\subparagraph{Weight states and ansatz class.} 
One of the working horses of our ansatz class will be the following family of unnormalized states.
Fix $a,\frakd,L$ and define
\begin{align}\label{def:WeigthStates}
  \ket{\chi_{a,\mathfrak{d},L}}=\sum\limits_{\substack{i_1+i_2+\ldots+i_L=a \\0\leq i_1,\ldots,i_L\leq \mathfrak{d}}}
  \ket{i_1i_2,\ldots,i_L} \in (\bbC^{\mathfrak{d}+1})^{\otimes L}.
\end{align}
We say that $\ket{\chi_{a,\mathfrak{d},L}}$ is the \emph{weight state} of weight $a$, length $L$ and local dimension $\frakd+1$. These weight states play a role in the differential geometry of homogeneous spaces (see, e.g., \cite[Chapter 12]{Ivey_2003}) and are of weight zero under certain actions of $SL_2$ (see, e.g., \cite{Pro:RepSL2Sperner}). Note that $L^{-\frac{1}{2}}\ket{\chi_{1,1,L}}$ is the $W$-state on $L$ sites. 

Suppose $\ket{\psi}$ is a state satisfying $\bbond_G(\ket{\psi}) \leq D$ and the degeneration has approximation degree $a$ and local degree at most $\frakd$. Then there are augmented linear maps $\calB^v : (\bbC^D)^{\otimes k_v} \otimes \bbC^{\frakd +1} \to \bbC^d$ such that 
\[
 \ket{\psi} = \left(\textbigotimes_v \calB^v \right) \left[ \ket{\Omega_D^{(G)}} \otimes \ket{\chi_{a,\frakd, L}} \right].
\]
Indeed, let $(\calA^v(\eps) : v \in V)$ be the sequence of local maps realizing the degeneration for $\ket{\psi}$, i.e., $\ket{\psi} = \lim_{\eps \to 0} \eps^{-a} \tns^G(\calA(\eps))= \lim_{\eps \to 0} \eps^{-a}(\bigotimes_v \calA^v(\eps)) \ket{\Omega_D^{(G)}}$. 

Expanding the local maps in terms of $\epsilon$, one has $\mathcal{A}^v(\epsilon)=\sum_{\eta=0}^{\mathfrak{d}} \calA^{v,\eta} \eps^\eta$, where $\calA^{v,\eta} : (\bbC^D)^{\otimes k_v} \to \bbC^d$ are linear maps. Define the augmented $\calB^v : (\bbC^D)^{\otimes k_v} \otimes \bbC^{\frakd+1} \to \bbC^d$ by 
\begin{equation}\label{def:augmented maps}
 \calB^v = \textsum_\eta \calA^{v,\eta} \otimes \langle \eta|.
\end{equation}
Specializing~\cite[Remark 9]{Christandl2017_TR} to our setting, we have 
\begin{align*}
\eps^{-a} \tns^G(\calA(\eps)) = &\eps^{-a} (\textbigotimes_v \calA^v(\eps)) \ket{\Omega_D^{(G)} }\\ & =(\textbigotimes_v \calB^v) (\ket{ \Omega_D^{(G)}} \otimes \ket{\chi_{a,\frakd,L}}) +\ket{Z(\eps)},
\end{align*}
where $\ket{Z(\eps)}$ converges to $0$ as $\eps \to 0$; in particular $\ket{\psi} = (\bigotimes_v \calB^v) (\ket{ \Omega_D^{(G)}} \otimes \ket{\chi_{a,\frakd,L}} )$. In other words, this construction results in only the term of $\ket{\psi(\eps)} = (\bigotimes_v \calA^v(\eps)) \ket{\Omega_D^{(G)} }$ of degree $a$ in $\eps$. 

\begin{figure}[ht!]
\includegraphics[width=0.95\columnwidth]{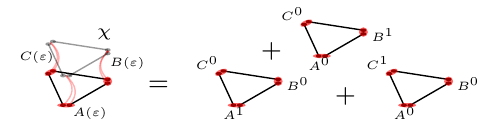}
\caption{Expansion of the TNS when contracted with the $\ket{\chi_{1,1,3}}$ state in terms of the local tensors corresponding to different degrees for $L=3$ and $a,\mathfrak{d}=1$.}
\label{fig:expansion_terms}
\end{figure}

This is illustrated in Fig.~\ref{fig:expansion_terms} for $a=1$ and $L=3$. Note that the pattern of the superpositions is mirrored in the entries of the weight state.

Given a family of linear maps $\calB = (\calB^v : (\bbC^D)^{\otimes k_v} \otimes \bbC^{\frakd+1} \to \bbC^d : v \in V)$, write 
\[
 \btns^G ( \calB)= \bigl( \textbigotimes_v \calB^v \bigr) \bigl( \ket {\Omega_D^{(G)}} \otimes \ket{\chi_{a,\frakd,L}}\bigr) 
\]
and define the set 
\[
\bTNS^G_{D,a,\frakd,d} = \bigl\{ \ket{\psi} = \btns^G(\calB) : \calB = (\calB^v : v \in V) \bigr\}.
\]
The parameters $a$ and $\mathfrak{d}$ should be regarded as additional variational parameters that play a similar role to the bond dimension $D$, in the sense that higher $a$ and $\mathfrak{d}$ increase the expressive power of the class, but also the complexity of optimizing over it. 

The discussion above implies that all states $\ket{\psi}$ satisfying $\bbond_G(\ket{\psi}) \leq D$ realized by a degeneration of approximation degree $a$ and local maps of degree at most $\frakd$ are contained in $\bTNS^G_{D,a,\frakd,d}$. However, $\bTNS^G_{D,a,\frakd,d}$ contains states that do not necessarily arise as a degeneration. Indeed, the family of maps defined in \eqref{def:augmented maps} satisfies relations which ensure the lower terms (in $\eps$) of the degeneration vanish. On the other hand, $\bTNS^G_{D,a,\frakd,d}$ is defined using arbitrary families of local maps $(\calB^v : v\in V)$: as a result, it contains all states that arise as structured superpositions mirroring the entries of the weight states. 

Optimizing solely over degenerations would entail optimizing over a subset of tensor network states satisfying $\cO(L^{a})$ \emph{global} polynomial equations of degree at most $a$ which encode the conditions that all terms of order strictly smaller than $a$ must vanish. Even just deciding whether the $0$-th order term vanishes in settings like PEPS on a square lattice is known to be an NP-complete problem~\cite{scarpa2018computational,tns_zero_testing}. Thus $\bTNS^G_{D,a,\frakd,d}$ is a superset of $\TNS^G_{D,d}$ which also contains states on the boundary and on which it is possible to optimize with a small overhead when compared with standard tensor network methods and without having to impose global equations, as we will show in the following.

From a numerical point of view, following the standard tensor network methods, one is interested in parameterizing the ansatz class in terms of the local maps. The space of such maps has dimension $d\sum_v D^{k_v}$ in the standard setting of $\TNS^G_{D,d}$ and has dimension $d(\frakd +1)\sum_v D^{k_v}$ in the setting of $\bTNS^G_{D,a,\frakd,d}$.

\subparagraph{Performing computations in the ansatz class.} 
An important feature of variational methods in the standard tensor network ansatz class is the possibility of computing expectation values of local observables  for tensors in the class. 

We evaluate the overhead to compute the expectation value if $\ket {\psi}$ has a representation in $\bTNS^G_{D,a,\frakd,d}$ compared to the case where it has a representation in $\TNS^G_{D,d}$.

A standard interpolation argument, see e.g. \cite{Bini:RelationsExactApproxBilAlg,Str:RelativeBilComplMatMult}, shows that if $\ket{\psi}$ is a state in $\cTNS^G_{D,d}$ and it is realized via a degeneration of error degree $e$, then $\ket{\psi} \in \TNS^G_{D(e+1),d}$. This fact is used in \cite{christandl2018tensor} to propose a contraction technique for states in $\cTNS^G_{D,d}$; however, even with constant local degrees $\frakd$, the error degree $e$ depends linearly in the system size $L$, therefore the complexity of this technique grows with $L$.

Here, we present two methods, which we call the \emph{MPS strategy} and the \emph{border rank strategy}, to perform the same type of computation in $\bTNS^G_{D,a,\frakd,d}$ more efficiently. These methods apply to states of $\bTNS^G_{D,a,\frakd,d}$ even if they are not elements of $\cTNS^G_{D,d}$; the additional information that the state arises as a degeneration, only provides a polynomial speed up in \cite{christandl2018tensor}. The MPS strategy consists of tensoring the standard tensor network representation of $\ket{\Omega_D^{(G)}}$ with an MPS representation of the desired weight state $\ket{\chi_{L,a,\frakd}}$ resulting in a tensor network representation of $\ket {\Omega_D^{(G)}} \otimes \ket{\chi_{L,a,\frakd}}$. The border rank strategy relies on the fact that the weight states have low \emph{border rank} \cite{Bini_1980,christandl2018tensor}, a semicontinuous version of tensor rank which is discussed in detail~\cite[Section I]{supplemental}. Which contraction technique is more advantageous is a subtle question, and highly depends on the combinatorics of the underlying graph. The MPS strategy allows for taking the geometry of the graph and the contraction order into account. For relevant cases like PEPS on a two-dimensional lattice, this contraction method provides an overhead bounded polynomially in $a$ (and in particular independent from $L$) in the contraction complexity when compared to contracting a PEPS of the same bond dimension. In contrast, the border rank strategy is oblivious to the geometry of graph and the contraction order, but the overhead in the contraction complexity is bounded by a $\cO((a+1)^2L)$ when compared to contracting a PEPS of the same bond dimension. We will now discuss the two strategies in more detail.

\begin{figure}[ht!]
\includegraphics[width=0.7\columnwidth,clip]{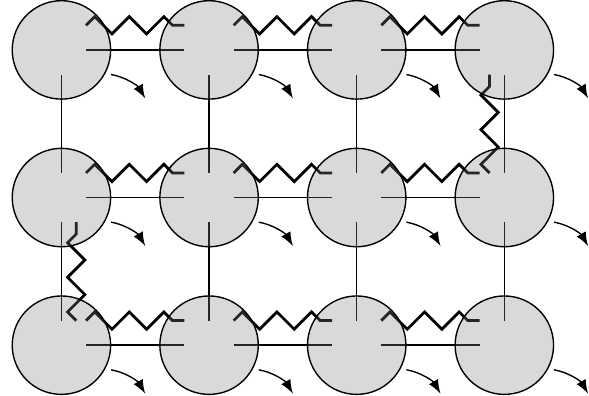}
\caption{PEPS on a $3\times 4$ square lattice $G$. The squiggly lines indicate the path $P$ supporting the MPS representation of $\ket{\chi_{a,\mathfrak{d},12}}$.}
\label{fig:peps_snake}
\end{figure}

For the MPS strategy, note that $\ket{\chi_{a,\mathfrak{d},L}}$ admits a representation as a matrix product state representation on an (open) chain of bond dimension $a+1$, see~\cite[Lemma 3]{supplemental}. Let $P$ be a path on $G$ which visits each vertex of $G$ at least once. One can ``lay'' the MPS representation of $\ket{\chi_{a,\mathfrak{d},L}}$ on the path $P$, resulting in a tensor network representation of $\Omega_D^{(G)} \otimes \ket{\chi_{a,\mathfrak{d},L}}$. The bonds of the resulting representation, however, are multiplied by a factor $(a+1)$ along each edge of $P$; in fact, the factor $(a+1)$ appears once for each time the corresponding edge appears in $P$; depending on the geometry of the graph, this might significantly increase the bonds of the final tensor network representation. In addition, this procedure is not translation invariant. However, in relevant cases, such as PEPS on a lattice, the path $P$ can be chosen so that it does not involve the same edge more than once: as a result, in this case, $\bTNS^{G}_{D,a,\frakd,d} \subseteq \TNS^G_{D(a+1)}$. In particular, in the case of the square lattice, contracting in $\TNS^G_{D(a+1)}$ has a complexity which is $(a+1)^4$ times the complexity of contracting on $\TNS^G_D$. Compared with the interpolation method proposed in~\cite{christandl2018tensor}, the MPS strategy proposed here is more efficient when $a$ is constant in the system size, as the complexity of the interpolation method scales with $L$. The MPS strategy is illustrated in Fig.~\ref{fig:peps_snake} for a square lattice.

As for the border rank strategy, ~\cite[Lemma 2]{supplemental} shows that the weight states admit expressions of the form $\ket{\chi_{a,\frakd,L}} = \lim_{\eps \to 0} \ket{\chi_{a,\frakd,L}(\eps)}$ where
\[
\ket{\chi_{a,\frakd,L}(\eps)} = \epsilon^{-a}\textsum_{i=1}^{a+1} \ket{x_i(\eps)}^{\otimes L} 
\]
is a sum of product vectors $\ket{x_i(\eps)}^{\otimes d}$ where $\ket{x_i(\eps)}$ is an element of $\bbC^{\frakd+1}$ depending linearly in $\eps$; in geometric language, this means that $\ket{\chi_{a,\frakd,L}}$ has border rank (at most) $a+1$. To get some intuition of why this is the case, note that we can also obtain the weight states as
\begin{align}\label{equ:representationderivative_main}
\ket{\chi_{a,a,L}}=a!\frac{d^a}{d \epsilon^a}\ket{\Gamma_a(\epsilon)}\big |_{\epsilon=0},
\end{align}
where
\begin{align*}
\ket{\Gamma_a(\epsilon)}=(\ket{0}+\epsilon \ket{1}+\ldots+\epsilon^{a}\ket{a})^{\otimes L}.
\end{align*}
Thus, expressing the derivative in \eqref{equ:representationderivative_main} as limit of a linear combination of $a+1$ points on the curve $\ket{\Gamma(\eps)}$, one obtains the claim. For example, if $a = \frakd = 1$, $\ket{\chi_{1,1,L}}$ is the unnormalized $W$-state on $L$ parties and one has $\ket{\chi_{1,1,L}} = \lim_{\eps \to 0} \left[(\ket{0}+\epsilon\ket{1})^{\otimes L}-\ket{0}^{\otimes L}\right]$.

This property allows one to compute expectation values using standard methods combined with an interpolation step. To see this, consider a converging sequence of states $\ket{\phi(\eps)}$ with $\ket{\phi} = \lim_{\eps \to 0}\ket{\phi(\eps)}$ and fix an observable $O$: we provide a technique to compute the expectation value $\langle \phi | O \phi\rangle$ assuming that we can only evaluate $\ket{\phi(\eps)}$ at nonzero values of $\eps$. As observed already in \cite{christandl2018tensor}, the function $p:\epsilon\mapsto p(\eps)=\eps^{-2a}\scalar{\phi(\bar{\eps})}{O\phi(\eps)}$ is a polynomial of degree at most $2(L\mathfrak{d}-a)$ in $\eps$ and its value at $\eps=0$ coincides with the desired expectation value. One cannot evaluate $p(\eps)$ at $\eps=0$ directly as the entries of the involved states diverge. However, via Lagrange interpolation, $p(0)$ is uniquely determined by the value of $p(\eps)$ at $2(L\mathfrak{d}-a)+1$ points. 

The border rank strategy applies this method to the function $p(\eps)$ when $\ket{\phi(\eps)} = \ket{\Omega^G_D }\otimes \ket{\chi_{a,\frakd,L}(\eps)}$. For instance, for $a=\frakd = 1$, $\ket{\chi_{1,1,L}(\epsilon)}$ is a superposition of two product states. We then have that $\ket{\phi(\epsilon)}$ can be written as the superposition of two tensor network states of bond dimension $D$. As a result, expanding $p(\eps)$ one sees that every single evaluation at $\eps \neq 0$ can be performed using standard tensor network methods.

We discuss this and the generalization to $a>1$ in more detail in~\cite[Section III]{supplemental}; we point out here that it is possible to evaluate the expectation value of any observable on $\ket{\phi}$ by contracting $(2(L\mathfrak{d}-a)+1)(a+1)^2$ tensor networks states of bond dimension $D$.

\subparagraph{Separations in the efficiency of representations.} It is natural to ask to what extent the $\bTNS^{G}_{D,a,\frakd,d}$ ansatz class provides more efficient representations of states of interest; in other words, we want to understand how large the gap between border bond dimension and bond dimension can be in the case of states of interest.

In the special case of matrix product states, i.e., when $G$ is a ring of length $L$,~\cite[Proposition 4]{supplemental} shows that $\cTNS_{D,d}^G \subseteq \TNS_{D^2,d}^G$; in other words, $\bbond^G(\ket{\psi}) \leq \bond^G(\ket{\psi}) \leq (\bbond^G(\ket{\psi})^2$. An analogous result holds for any graph, but the exponent depends on the combinatorics of the graph and in particular it may depend on $L$, making the upper bound exponential in the system size. For a general graph, as mentioned before, if $\bbond^G(\ket{\psi}) \leq D$ and the degeneration has error degree $e$, then $\bond^G(\ket{\psi}) \leq (e+1)D$. 

Little is known about lower bounds on the possible separation. The difficulty in obtaining examples of large separations between bond and border bond dimension lies in the fact that essentially all techniques to prove lower bounds for bond dimension give, in fact, a lower bound on the border bond dimension: 
this is the case for the rank across a cut and other methods relying on the evaluation of semicontinuous functions.

In the case where $G$ is the ring with three nodes, consider $\ket{\ghz_3}$ be the level three GHZ state on three parties; it has long been known \cite{Str:RelativeBilComplMatMult} that $\bbond^G(\ket{\ghz_3}) = 2$; in \cite{christandl2018tensor}, the authors show $\bond^G(\ket{\ghz_3}) = 3$, showing a separation. An additional example is provided \cite{christandl2018tensor}, where the possibility that the same separation holds also for the RVB state on the kagome lattice is discussed.

One can determine examples where the separation depends on the system size in the setting of translation invariant (TI) tensor networks. Consider translationally invariant matrix product states with periodic boundary conditions on an odd number of vertices $L$. Define 
\begin{align}\label{equ:definitionpsisep}
\ket{\psi}=\frac{1}{\sqrt{L}}\sum\limits_{k=0}^{L-1}S^k\ket{21010\ldots10} \in (\C^3)^{\otimes L},
\end{align}
where $S$ is the shift operator. 
Consider the projector $P$ acting on $\C^{3}\otimes \C^{3}$ given by:
\begin{align}\label{equ:definitonprojection}
P=\ketbra{01}{01}+\ketbra{10}{10}+\ketbra{02}{02}+\ketbra{21}{21}.
\end{align}
Define a Hamiltonian $H$ on a ring of size $L$ as:
\begin{align}\label{equ:definitionHamiltonian}
H=\sum_{i=0}^{L-1} \left[ (I-P_{i,i+1})+\textstyle\frac{1}{2L}\ketbra{2}{2}_i\right],
\end{align}
where $I$ is the identity map, $P_{i,i+1}$ acts as $P$ in Eq.~\eqref{equ:definitonprojection} on sites $i,i+1$ and addition is taken modulo $L$.
From~[Proposition 6]\cite{supplemental} we have that $\ket{\psi}$ is the unique translationally invariant ground state of this $2$\nobreakdash-local Hamiltonian on the ring of length $L$ for $L$ odd.

Moreover, note that $\bbond^{\text{TI-}G}(\ket{\psi}) = 2$ with a degeneration having $a = \frakd = 1$. Indeed, consider the degeneration defined by the local map $\calA(\eps) : \bbC^2 \otimes \bbC^2 \to \bbC^3$ defined as follows: write $\calA(\eps) = \calA^0 + \eps \calA^1$ with
\begin{equation}\label{equ:degenerationpsi}
\begin{aligned}
\calA^0 &= \ketbra{0}{1} \otimes \ket{0} + \ketbra{1}{0} \otimes \ket{1},\\
\calA^1 &= \ketbra{0}{0} \otimes \ket{2};
\end{aligned} 
\end{equation}

it is easy to see that $\ket{\psi} = \lim_{\eps \to 0} \epsilon^{-1}\calA(\eps)^{\otimes L} (\ket{\Omega_D^G})$. In particular, $\ket{\psi} \in \bTNS^{\text{TI-}G}_{2,1,1,3}$. On the other hand, an adaptation of the results of~\cite{mapsgarcia} and \cite{Michalek_2019_Wielandt} shows that $\bond^{\text{TI-}G}(\ket{\psi})= \Omega\lb L^{1/3} / \log(L)\rb$. This implies system-size dependent separations asymptotically for large enough $L$. Furthermore, the results of~\cite{mapsgarcia} also imply that $\bond^{\text{TI-}G}(\ket{\psi})>2$ for all ring sizes.

\subparagraph{Variational methods.} In this section, we discuss two widely used methods to find ground states of local Hamiltonians using tensor networks which can be adapted to the $\bTNS^{G}_{D,a,\frakd,d}$  ansatz class: gradient descent \cite{Pirvu2010},~\cite[Section 7.1]{Orus_practical} and imaginary time evolution ~\cite[Section 7.2]{Orus_practical}, which is sometimes called decimal block decimation method in this context. As in the case of the computation of expectation values, it is possible to adapt usual tensor network techniques and code to also optimize within the $\bTNS^{G}_{D,a,\frakd,d}$ class with minimal effort. 

First, we discuss gradient descent methods to find the ground state of a local Hamiltonian. In the standard tensor network setting, given a local Hamiltonian $H$ and a desired bond dimension, one considers the energy of a tensor network state $\tns^G(\calA)$ as a function depending on the family of local maps $\calA = (\calA^v: v \in V)$. A gradient method, computes the gradient of energy function and optimizes with respect to the linear maps. Often, it is useful to consider the energy function as a function of only one linear map, optimize with respect to that, and then repeating the procedure alternating among all the linear maps: this method is called \emph{alternating gradient descent} and we refer to \cite[Section 7.2]{Orus_practical} for more details. In the translation invariant case, one considers the energy as a function of a single linear map and optimizes with respect to that, as in \cite{Pirvu2010}. In the boundary setting, the same method can be used considering the energy of a state $\btns(\calB)$ in $\bTNS^G_{D,a,\frakd,d}$ as a function of the family of linear maps $\calB = (\mathcal{B}^{v_1},\ldots,\mathcal{V},\ldots,\mathcal{B}^{v_L})$, with $\calB^{v_j} : (\bbC^D)^{\otimes k_{v_j}} \otimes \bbC^{\frakd+1} \to \bbC^d$. 

We give some details to explain how to compute the gradient efficiently in the $\bTNS^{G}_{D,a,\frakd,d}$ class. Suppose we are computing the directional derivative of the function $\langle \btns(\mathcal{B})| O \ \btns(\mathcal{B}) \rangle$  for some observable $O$ in the direction of a vector $\mathcal{V}$ on the $k$-th component of $\calB$. Then
\begin{align*}
&\frac{\partial}{\partial \mathcal{V}} \langle \btns(\mathcal{B})| O \ \btns(\mathcal{B}) \rangle=\\
&2\textrm{Re } \langle \btns(\mathcal{B}^{v_1},\ldots,\mathcal{B}^{v_{k-1}},\mathcal{V},\mathcal{B}^{v_{k+1}},\ldots,\mathcal{B}^{v_L})| O \ \btns(\mathcal{B}) \rangle,
\end{align*}
which is the overlap of two states in $\bTNS^G_{D,a,\frakd,d}$ with respect to the observable $O$. As explained before, the calculation of the overlap can be done by combining standard contraction methods for tensor network states and one of the contraction strategies outlined before for states in $\bTNS^G_{D,a,\frakd,d}$. In order to optimize the energy of some local Hamiltonian by alternating gradient descent, one computes the partial derivatives of the function
\begin{align*}
   f(\mathcal{B})=\frac{E(\mathcal{B})}{N(\mathcal{B})}
\end{align*}
where $E(\mathcal{B})=\langle \btns(\mathcal{B})| H \ \btns(\mathcal{B}) \rangle$ and $N(\mathcal{B})=\langle \btns(\mathcal{B})| \btns(\mathcal{B})\rangle$. The calculation of the directional derivatives of $f$ reduces to the valuation of $E,N$ and of the their derivatives, which can be done as explained above. Note that this method can be easily generalized to perform gradient descent in order to maximize the overlap with another state.

The second variational method that we consider is \emph{imaginary time evolution}. Imaginary time evolution relies on the fact that, given a local Hamiltonian $H$ and a state $\ket{\psi}$ with nonzero overlap with the ground state, the state $e^{-\beta H}\ket{\psi}/\|e^{-\beta H}\ket{\psi}\|$ converges to the ground state of $H$ as $\beta$ diverges to infinity. The method approximates the map $e^{-\beta H}$ by a sequence of local maps through Trotterization. This sequence of local maps is applied to the current state, and it is easy to see that the resulting state has a larger overlap with the ground state than the initial one. This procedure can be easily implemented in $\bTNS^G_{D,a,\frakd,d}$ by applying the local map to the physical indices of the underlying tensors. 

More precisely, let $e^{-\beta H_{v_1v_2}}:\lb\C^d\rb^{\otimes 2}\to\lb\C^d\rb^{\otimes 2}$ be a two-local imaginary time evolution term acting on nodes $v_1$ and $v_2$ connected by an edge $e_1$. Consider a Schmidt decomposition of the operator $e^{-\beta H_{v_1v_2}}$:
\begin{align*}
  e^{-\beta H_{v_1v_2}}=\sum_{\ell=1}^{d^2}X_{\ell}\otimes Y_{\ell} .
\end{align*}
Then the vector $e^{-\beta H_{v_1v_2}}\btns^G ( \calB)$ can be obtained directly by enlarging the bond dimension across the edge $e_1$ by $d^2$ and updating the local maps $\mathcal{B}^{v_1}$ and $\mathcal{B}^{v_2}$. Thus, after applying one two-local map on each edge of the network, a state in $\bTNS^G_{D,a,\frakd,d}$ is mapped to a state in $\bTNS^G_{d^2D,a,\frakd,d}$. Hence, the bond dimension will (potentially) increase exponentially with the number of applied steps.

In the standard tensor network setting, this is addressed by truncating the Schmidt decomposition of $e^{-\beta H_{v_1v_2}}\tns^G ( \calA)$. In the following, we show that the $\bTNS^{G}_{D,a,\frakd,d}$ ansatz class supports a suitable truncation of the bond dimension after a certain number of iterations; in other words, we provide a method of finding an approximation of a state $\ket{\psi_1} \in \bTNS^G_{D_1,a,\frakd,d}$ by a state $\ket{\psi_2} \in \bTNS^G_{D_2,a,\frakd,d}$ with $D_2<D_1$. The gradient descent methods discussed before can be used to this end, as we can optimize the overlap of the state $\ket{\psi_1}$ with respect to states in $\bTNS^G_{D_2,a,\frakd,d}$. However, a method to perform this truncation by solely considering the local maps $\mathcal{B}^v$ is desirable, as in every iteration of gradient descent the whole state has to be contracted.

In the case of MPS with open boundary conditions, this problem is solved by first contracting two subsequent local maps $\mathcal{A}^{v_1},\mathcal{A}^{v_2}:\C^D\otimes \C^D\to\C^d$ along their common edge. Let $\mathcal{A}^{v_1 v_2}$ be the new map we obtain this way. Seeing it as matrix $\mathcal{A}^{v_1 v_2}:\C^d\otimes \C^D\to\C^d\otimes \C^D$, we then perform a singular value decomposition and discard all singular values below a certain threshold. After the truncation, we obtain new local maps $\mathcal{A}^{v_1},\mathcal{A}^{v_2}$ with a smaller bond dimension on that edge, as desired. A Schmidt decomposition shows that this form of truncation is indeed optimal. 
However, in the case of tensor networks with cycles, the optimal truncation strategy is a subtle issue~\cite{Evenbly_2018}. This is primarily due to the fact that edges on a cycle do not induce a bipartition of the state and, thus, a Schmidt decomposition of the state. Nevertheless, truncations based on a purely local SVD truncation, the simple-update algorithm~\cite{Jiang_2008}, perform well in practice.

In principle, SVD-based truncation techniques can be readily applied to the $\bTNS^G_{D,a,\frakd,d}$ class.
We can simply again contract $\calB^{v_1}$ and $\calB^{v_2}$ along the shared edge, obtaining a map $\calB^{v_1v_2}$. We then perform a SVD and subsequent truncation of the matrix $\calB^{v_1v_2}:\C^d\otimes\C^{\frakd}\otimes \lb\C^{D}\rb^{\otimes (k_{v_1}-1)}\to \C^d\otimes\C^{\frakd}\otimes \lb\C^{D}\rb^{\otimes (k_{v_1}-1)}$.

This would allow us to truncate the bond dimension of states in $\bTNS^G_{D,a,\frakd,d}$ locally. However, note that this truncation strategy did not take the special structure of the states in $\bTNS^G_{D,a,\frakd,d}$ into account, which may lead to  suboptimal truncations. To illustrate this more concretely, take $a,\mathfrak{d}=1$ and consider the representation of the state $\ket{\psi}$ in eq.~\eqref{equ:degenerationpsi} with bond dimension $4$ given with the same local tensors up to $\mathcal{A}^1$, which we now set to
\begin{align*}
\mathcal{A}^1&=\ketbra{0}{0}\otimes \ket{2}+2\ketbra{2}{2}\otimes \ket{2}\\&+2\ketbra{3}{3}\otimes \ket{2}.
\end{align*}
Note that adding this extra subspace did not change the resulting state. The matrix $\mathcal{B}^{v_1v2}$ we obtain for this representation is:
\begin{align}\label{equ:matrixc}
 \mathcal{B}^{v_1v_2}=\left(\begin{array}{ccc|ccc}0 &\ketbra{0}{0}& 0 &0 &0 &0 \\
\ketbra{1}{1} & 0& 0 &0 &0 & \ketbra{1}{0}\\
0 & 0& 0 &0 &0 & 0\\
\hline
0 & 0& 0 &0 &0 & 0\\
0 & 0& 0 &0 &0 & 0\\
0 & \ketbra{0}{1} &0 &0 &0 & \mathcal{A}_2^\dagger\mathcal{A}_2
\end{array}\right),
\end{align}
where each entry corresponds to a $4\times 4$ matrix. Note that the upper left $3\times 3$ block submatrix corresponds to the crossing of degree $0$ terms, while the upper right and lower left correspond to degree $0$ and $1$ terms. Finally, the lower right corresponds to the crossing of degree $1$ with $1$.
Observe that the lower right submatrix does not contribute to the resulting state when we contract with the state $\ket{\chi_{1,1,L}}$. This is because it corresponds to a term of degree $2$. Thus, it is important to take this into account when performing a truncation.

The matrix in~\eqref{equ:matrixc} is an extreme example of the issue that not all parts of the submatrix $C$ contribute equally to the state. Suppose that we wish to truncate this bond from three to two by performing a SVD of~\eqref{equ:matrixc} and discarding the four smallest singular values. This truncation would lead  us to discard the subspace spanned by $\ket{0},\ket{1}$. But this choice of truncation would then result in the $0$ state, as it only preserved the submatrix  corresponding to degree $2$ and set the other to $0$.

We see we should not care if a truncation changes the lower right matrix substantially, while approximately preserving the other blocks. Furthermore, the upper right and lower left matrix contribute only once to the resulting state, while the submatrices corresponding to constant order degrees appear $L-1$ times.

In light of this, let us now discuss an heuristic algorithm to perform truncations taking the special structure of the states in $\bTNS^G_{D,a,\frakd,d}$ into account.
The first step is to recall the variational formulation of the truncation of singular values. For a matrix $C$ with SVD $C=UDV$ and some truncation rank $r$, denote by $\sqrt{D_r}$ a $n\times r$ matrix containing the square root of the largest $r$ singular values of $C$ on the diagonal. It is well-known that $A=U\sqrt{D_r}$, $B=\sqrt{D_r}^{T}V$ minimize 
\begin{align}\label{equ:variationalsvd}
    (A_r,B_r)\mapsto\|C-A_rB_r\|_{F},
\end{align}
where $A_r$ and $B_r$ are rank $r$ matrices and $\|\cdot\|_{F}$ is the Frobenius norm. This is sometimes referred to as the Eckart-Young theorem.
Our approach will be based on picking a different norm to perform this optimization depending on a parameter $0<p\leq1$. This parameter $p$ encodes by how much we want to suppress submatrices that correspond to higher orders. For a map $\calB^{v_1 v_2}:\C^{\mathfrak{d}}\otimes \C^d \otimes \lb \C^{D}\rb^{\otimes l} :\C^{\mathfrak{d}}\otimes \C^d 
\otimes \lb \C^{D}\rb^{\otimes l}$ consider the weighted Frobenius norm:
\begin{align*}
    \|\calB^{v_1 v_2}\|_{F,p}^2=\sum\limits_{\eta_1,\eta_2=0}^{\mathfrak{d}}p^{2\lb\eta_A+\eta_B\rb}\sum\limits_{i,j,\alpha,\beta}\left|\calB^{v_1 v_2}_{(\eta_1,i,\alpha),(\eta_2,j,\beta)}\right|^2.
\end{align*}
We see that submatrices that correspond to higher degrees contribute less to this norm. Indeed, if we denote by $\calB^{v_1 v_2}_{\eta}$ the submatrix for indices such that $\eta_1+\eta_2=\eta$ we have that
\begin{align*}
\|\calB^{v_1 v_2}-\tilde{\calB}^{v_1 v_2}\|_{F,p}^2=\sum\limits_{\eta=0}^{\mathfrak{d}}p^{2\eta}\|\calB^{v_1 v_2}_\eta-\tilde{\calB}^{v_1 v_2}_\eta\|_{F}^2
\end{align*}
for any two matrices $\calB^{v_1 v_2}_{\eta},\tilde{\calB}^{v_1 v_2}_{\eta}$. 
In the previous example with $a,\mathfrak{d}=1$, the Frobenius norm of the upper left matrix is multiplied by $1$, the upper right and lower left by $p$ and the lower right by $p^2$.
We see that this norm is less sensitive to the Frobenius distance of the submatrices corresponding to higher degree terms. Thus, truncating the bond dimension w.r.t. to this norm, truncation errors in higher degree terms will contribute less, as desired, and the parameter $p$ controls by how much. For example, in the case of $\calB^{v_1 v_2}$ in eq.~\eqref{equ:matrixc}, we see that picking $p< 1/2$ is enough to ensure that we truncate the subspace spanned by $\ket{2},\ket{3}$ is discarded, as desired. 
Unfortunately, it is not clear at  this point how to pick the parameter $p$ in an optimal fashion. One possibility is to perform the truncation for different values of $p$ and compare the resulting overlap with the original state.

Moreover, performing the truncation w.r.t. this norm can be easily implemented through standard SVD techniques combined with a rescaling step. Defining the new map $\tilde{\calB}^{v_1 v_2}$ with entries 
\begin{align*}
\tilde{\calB}^{v_1 v_2}_{(\eta_1,i,\alpha),(\eta_2,j,\beta)}=p^{\eta_1+\eta_2}\calB^{v_1 v_2}_{(\eta_1,i,\alpha),(\eta_2,j,\beta)},
\end{align*}
it is easy to see that $\|\tilde{\calB}^{v_1 v_2}\|_{F}=\|\calB^{v_1 v_2}\|_{F,p}$. 

Let $\tilde{\calA}^{v_1}_r$ and $\tilde{\calA}^{v_2}_r$ be the matrices obtained by performing a truncated SVD of $\tilde{\calB}^{v_1 v_2}$ to rank $r$, where $r$ is the desired truncated bond dimension. By the Eckart-Young theorem, we have that $\tilde{\calA}^{v_1}_r$, $\tilde{\calA}^{v_2}_r$ minimize $\|\tilde{\calB}^{v_1 v_2}-\tilde{\calA}^{v_1}_r\tilde{\calA}^{v_2}_r\|_{F}$ amongst all matrices of rank at most $r$.
It is then easy to see that by defining $\calA_{(\eta_1,i,\tilde{\alpha})}^{v_1}=p^{-\eta_1}\tilde{\calA}_{(\eta_1,i,\tilde{\alpha})}^{v_1}$ and $\calA_{(\eta_1,i,\tilde{\alpha})}^{v_2}$ analogously, we have that $\calA^{v_1},\calA^{v_2}$ minimize $\|\calA^{v_1v_2}-\calA^{v_1}\calA^{v_2}\|_{F,p}$, in the spirit of Eq.~\eqref{equ:variationalsvd}. 

Thus, by suitably re-scaling the initial tensor, performing the usual truncation methods through an SVD and then scaling back, it is possible to perform the truncation taking into account the contribution of each subspace and only performing local operations. Moreover, the computational complexity of performing all of these steps is comparable with that of locally truncating the bond dimension of a state in the tensor network setting, as they only differ by the rescaling steps.

\subparagraph{Numerical results.}
In this section, we compare numerical methods in the standard tensor network setting with the analogous methods in the $\bTNS^{G}_{D,a,\frakd,d}$ ansatz class. Let us start by illustrating how we expect numerics to behave when trying to approximate a state on the boundary through states in the interior in an example. To this end, consider the state $\ket{T}\in (\C^9)^{\otimes 3}$ given by
\begin{align}\label{equ:defiTstate}
  &\ket{T}=\frac{1}{\sqrt{17}}  (\ket{005}+\ket{016}+\ket{040}+\ket{126}\\\nonumber
  &+\ket{160}+\ket{227}+\ket{251}+\ket{262}+\ket{338}+\ket{373}\\\nonumber
  &+\ket{384}+\ket{430}+\ket{501}+\ket{632}+\ket{703}+\ket{714}\\\nonumber
  &+\ket{824}).\nonumber
\end{align}
In~\cite[Section II.B]{supplemental}, we show that $\ket{T}$ satisfies
\begin{align*}
    3=\bbond^{C_3}(\ket{T})<\bond^{C_3}(\ket{T}),
\end{align*}
where $C_3$ is the ring on three vertices. Moreover, the approximation degree of the border bond dimension representation is $a=1$.

\begin{figure}[ht!]
\includegraphics[width=0.9\columnwidth,trim={0.1cm 0cm 0 0cm},clip]{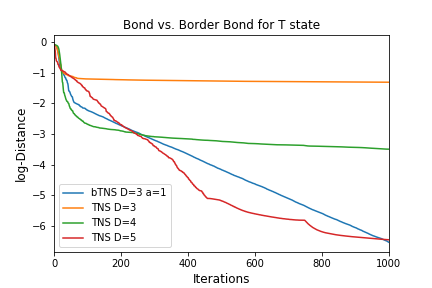}
\caption{Logarithm of the distance to $\ket{T}$ after number of iterations for alternating gradient descent and different bond dimensions on the ring.}
\label{fig:T}
\end{figure}

Fig~\ref{fig:T} records the results of performing alternating gradient descent to approximate the state $\ket{T}$ in different ansatz classes. More precisely, we are performing alternating gradient descent by optimizing the overlap with the bond dimension $17$ representation of the state $\ket{T}$ given by the decomposition in Eq.~\eqref{equ:defiTstate}. We do so because we are unaware of a better representation of $\ket{T}$, although~\cite[Lemma 5]{supplemental} guarantees that $\bond^{C_3}(\ket{T}) \leq 9$.
Although the state $\ket{T}$ can be approximated arbitrarily well with states of bond dimension $3$, we observe that the convergence is slow. Although by ~\cite[Lemma 5]{supplemental}  the maximal bond dimension is $D=9$, we already observe a fast convergence with $D=5$. On the other hand, since $\bbond^{C_3}(\ket{T}) = 3$ with a degeneration having approximation degree $1$, we have $\ket{T} \in \bTNS^{C_L}_{3,1,1,9}$: indeed performing gradient descent in this class we already observe fast convergence, which can only observed for bond dimension $5$ in the standard tensor network setting. 

\begin{figure}[ht!]
\includegraphics[width=0.9\columnwidth]{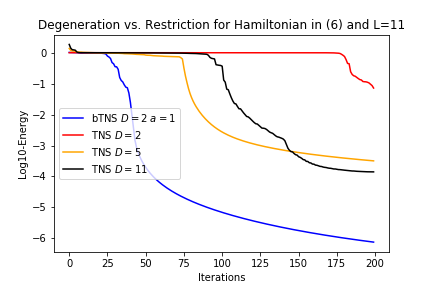}
\caption{Energy achieved for a TI-TNS ansatz compared to bTNS for the Hamiltonian defined in eq.~\eqref{equ:definitionHamiltonian}. The ground state energy is normalized to $0$ and the ring is of size $11$. Note that the time taken for one iteration of bTNS for $D=2$ is roughly half that of one in TNS with $D=5$. We picked the best energy value achieved over 40 random starting points for each curve. We note a faster convergence for bTNS.}
\label{fig:ourhamiltonian}
\end{figure}

Fig.~\ref{fig:ourhamiltonian} records the results in the translation invariant setting of the Hamiltonian described in \eqref{equ:definitionHamiltonian}. Fix a ring of length $L = 9$. The ground state $\ket{\psi}$ of \eqref{equ:definitionpsisep} satisfies $\bbond^{C_{11}-TI} (\ket{\psi}) = 2$ with approximation degree $a=1$, whereas $\bbond^{C_{11}-TI} (\ket{\psi}) >2$. Gradient descent on $\bTNS^{C_{11}-TI}_{2,1,1,3}$ outperforms the standard matrix product state ansatz for $D=2,4$, in the sense that it obtains lower energies, giving an indication of the numerical viability of our method. Moreover, an iteration of gradient descent with $D=4$ in $\TNS$ takes roughly double the time of one in $\bTNS$ with $D=2$ and $a=1$.

\begin{figure}[ht!]
\includegraphics[width=0.9\columnwidth,,clip]{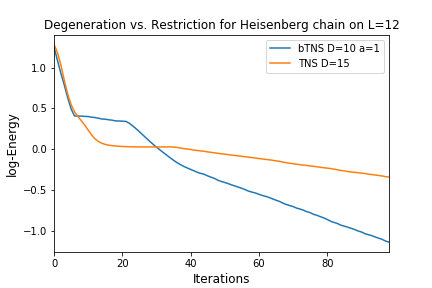}
\caption{Energy achieved with the imaginary time evolution for the isotropic Heisenberg model for a translation invariant ansatz and $L=12$. We normalized the ground state energy to be $0$ and picked $p=0.9$ for the truncation algorithm of the degeneration. We observe a faster convergence for bTNS.}
\label{fig:heisenbergmodel}
\end{figure}

In order to benchmark the $\bTNS^{G}_{D,a,\frakd,d}$ ansatz class and the algorithms for imaginary time evolution, we performed the method on the isotropic Heisenberg model on a ring of size $L$, comparing the results achieved by the standard matrix product state method with the ones obtained in the $\bTNS^{G}_{D,a,\frakd,d}$ ansatz class. The Hamiltonian of the isotropic Heisenberg model on the ring $C_L$ is given by
\begin{align*}
    H=\sum_{k=1}^{L} \lb \sigma_{k}^x\sigma_{k+1}^x+\sigma_{k}^y\sigma_{k+1}^y+\sigma_{k}^z\sigma_{k+1}^z\rb
\end{align*}
where $\sigma_{k}^i$ are the Pauli matrices acting on site $k$.
The number of local parameters for a state in our ansatz class with bond dimension $D$ is $(a+1)D^2$, so we compare states in our class to MPS with bond dimension $\lceil \sqrt{a+1}D\rceil$. 
We used imaginary time evolution methods and a translationally invariant ansatz to find the ground state and picked the initial tensor at random.
We see in Fig.~\ref{fig:heisenbergmodel} that states in our class converge faster. Although we do not have a provable separation in the required bond dimension for this model, these results indicate the potential of our method for models of physical interest.

\subparagraph{Conclusion.}
We presented numerical evidence showing that degenerations of tensor network states are a valuable tool for both the numerical and analytical study of the tensor network ansatz class. Many directions remain to be explored in future work, both from the analytical and numerical point of view. On the numerical side, it will be interesting to go beyond one dimensional systems and see how our enlarged ansatz class performs for higher dimensional lattices. In particular, we believe that larger separations in complexity can be observed for PEPS, even when not restricting to translationally invariant systems. Moreover, we believe that our ansatz class provides a natural framework to study excited states.
From the analytical and algorithmic point of view, our work raises many questions. Just to name a few, it is natural to ask about normal forms, the scaling of correlations and how degenerations behave in the thermodynamical limit.

This work was supported by VILLUM FONDEN via the QMATH Centre of Excellence under Grant No. 10059 and the European Research Council (Grant agreement No. 818761). AHW thanks the VILLUM FONDEN for its support with a Villum Young Investigator Grant (Grant No. 25452).

\nocite{BalBerChrGes:PartiallySymRkW,efremenko94barriers,Hackbusch_2019,Haferkamp_2020,hoorfar2008inequalities,Lan:TensorBook,Quarteroni_2006,Rahaman_2020,sanz2010quantum,PhysRevLett.98.140506,Sylv:PrinciplesCalculusForms,Verstraete_2006}

\bibliographystyle{plain}
\bibliography{ansatzclass.bib}

\begin{thebibliography}{10}

\bibitem{supplemental}
See the Supplemental Material at [insert url] for details on border rank and
  MPS representations of the weight states, separation between bond and border
  bond dimension, approximations of states on the boundary of the tensor
  network variety by ones in the interior and stable polynomial interpolation.

\bibitem{Arad_2017}
I.~Arad, Z.~Landau, U.~Vazirani, and T.~Vidick.
\newblock {Rigorous {RG} Algorithms and Area Laws for Low Energy Eigenstates in
  1D}.
\newblock {\em Communications in Mathematical Physics}, 356(1):65--105, aug
  2017.

\bibitem{BalBerChrGes:PartiallySymRkW}
E.~Ballico, A.~Bernardi, M.~Christandl, and F.~Gesmundo.
\newblock {On the partially symmetric rank of tensor products of W-states and
  other symmetric tensors}.
\newblock {\em Atti Accad. Naz. Lincei Rend. Lincei Mat. Appl.}, 30:93--124,
  2019.

\bibitem{Bini:RelationsExactApproxBilAlg}
D.~Bini.
\newblock {Relations between exact and approximate bilinear algorithms.
  {A}pplications}.
\newblock {\em Calcolo}, 17(1):87--97, 1980.

\bibitem{Bini_1980}
D.~Bini, G.~Lotti, and F.~Romani.
\newblock {Approximate Solutions for the Bilinear Form Computational Problem}.
\newblock {\em {SIAM} Journal on Computing}, 9(4):692--697, nov 1980.

\bibitem{ChrGesJen:BorderRankNonMult}
M.~Christandl, F.~Gesmundo, and A.~K. Jensen.
\newblock {Border rank is not multiplicative under the tensor product}.
\newblock {\em SIAM J. Appl. Alg. Geom.}, 3:231--255, 2019.

\bibitem{Christandl2017_TR}
M.~Christandl, A.~K. Jensen, and J.~Zuiddam.
\newblock {Tensor rank is not multiplicative under the tensor product}.
\newblock {\em Linear Algebra and its Applications}, 543:125--139, 2018.

\bibitem{christandl2018tensor}
M.~Christandl, A.~Lucia, P.~Vrana, and A.~H. Werner.
\newblock {Tensor network representations from the geometry of entangled
  states}.
\newblock {\em SciPost Phys.}, 9(3):42, 2020.

\bibitem{efremenko94barriers}
K.~Efremenko, A.~Garg, R.~Oliveira, and A.~Wigderson.
\newblock {Barriers for Rank Methods in Arithmetic Complexity}.
\newblock In {\em {9th Innovations in Theoretical Computer Science Conference
  (ITCS 2018)}}. Schloss Dagstuhl-Leibniz-Zentrum fuer Informatik, 2018.

\bibitem{Evenbly_2018}
G.~Evenbly.
\newblock {Gauge fixing, canonical forms, and optimal truncations in tensor
  networks with closed loops}.
\newblock {\em Physical Review B}, 98(8), aug 2018.

\bibitem{Fannes_1992}
M.~Fannes, B.~Nachtergaele, and R.~F. Werner.
\newblock {Finitely correlated states on quantum spin chains}.
\newblock {\em Communications in Mathematical Physics}, 144(3):443--490, mar
  1992.

\bibitem{tns_zero_testing}
S.~Gharibian, Z.~Landau, S.~Shin, and G.~Wang.
\newblock {Tensor network non-zero testing}.
\newblock {\em Quantum Information and Computation}, 15, 06 2014.

\bibitem{Hackbusch_2019}
W.~Hackbusch.
\newblock {A {Note} on {Nonclosed} {Tensor} {Formats}}.
\newblock {\em Vietnam Journal of Mathematics}, 48(4):621--631, December 2020.

\bibitem{Haferkamp_2020}
J.~Haferkamp, D.~Hangleiter, J.~Eisert, and M.~Gluza.
\newblock {Contracting projected entangled pair states is average-case hard}.
\newblock {\em Physical Review Research}, 2(1), jan 2020.

\bibitem{Hauschild_2018}
J.~Hauschild and F.~Pollmann.
\newblock {Efficient numerical simulations with Tensor Networks: Tensor Network
  Python ({TeNPy})}.
\newblock {\em {SciPost} Physics Lecture Notes}, oct 2018.

\bibitem{hoorfar2008inequalities}
A.~Hoorfar and M.~Hassani.
\newblock {Inequalities on the Lambert W function and hyperpower function}.
\newblock {\em J. Inequal. Pure and Appl. Math}, 9(2):5--9, 2008.

\bibitem{Ivey_2003}
T.~Ivey and J.~M. Landsberg.
\newblock {\em {Cartan for Beginners}}.
\newblock American Mathematical Society, sep 2003.

\bibitem{Jiang_2008}
H.~C. Jiang, Z.~Y. Weng, and T.~Xiang.
\newblock {Accurate Determination of Tensor Network State of Quantum Lattice
  Models in Two Dimensions}.
\newblock {\em Physical Review Letters}, 101(9), aug 2008.

\bibitem{Landau_2015}
Z.~Landau, U.~Vazirani, and T.~Vidick.
\newblock {A polynomial time algorithm for the ground state of one-dimensional
  gapped local Hamiltonians}.
\newblock {\em Nature Physics}, 11(7):566--569, jun 2015.

\bibitem{Lan:TensorBook}
J.~M. Landsberg.
\newblock {\em {Tensors: {G}eometry and {A}pplications}}, volume 128 of {\em
  {Graduate Studies in Mathematics}}.
\newblock American Mathematical Society, Providence, RI, 2012.

\bibitem{Landsberg_2017}
J.~M. Landsberg.
\newblock {\em {Geometry and Complexity Theory}}.
\newblock Cambridge University Press, 2017.

\bibitem{Landsberg_geometry}
J.~M. Landsberg, Y.~Qi, and K.~Ye.
\newblock {On the Geometry of Tensor Network States}.
\newblock {\em Quantum Info. Comput.}, 12(3--4):346--354, March 2012.

\bibitem{Michalek_2019_Wielandt}
M.~Michalek and Y.~Shitov.
\newblock {Quantum Version of Wielandt's Inequality Revisited}.
\newblock {\em {IEEE} Transactions on Information Theory}, 65(8):5239--5242,
  aug 2019.

\bibitem{Orus_practical}
R.~Or{\'u}s.
\newblock {A practical introduction to tensor networks: Matrix product states
  and projected entangled pair states}.
\newblock {\em Annals of Physics}, 349:117--158, oct 2014.

\bibitem{mapsgarcia}
D.~Perez-Garcia, F.~Verstraete, M.~M. Wolf, and J.~I. Cirac.
\newblock {Matrix Product State Representations}.
\newblock {\em Quantum Info. Comput.}, 7(5):401--430, July 2007.

\bibitem{Pirvu2010}
B.~Pirvu, F.~Verstraete, and G.~Vidal.
\newblock {Exploiting translational invariance in matrix product state
  simulations of spin chains with periodic boundary conditions}.
\newblock {\em Physical Review B}, 83(12), mar 2011.

\bibitem{Pro:RepSL2Sperner}
R.~A. Proctor.
\newblock {Representations of $\mathfrak{sl}(2,\mathbb{C})$ on Posets and the
  Sperner Property}.
\newblock {\em SIAM J. Alg. Disc. Methods}, 3(2):275--280, 1982.

\bibitem{Quarteroni_2006}
A.~Quarteroni, R.~Sacco, and F.~Saleri.
\newblock {\em {Numerical Mathematics}}.
\newblock Springer New York, 2007.

\bibitem{Rahaman_2020}
M.~Rahaman.
\newblock {A New Bound on Quantum Wielandt Inequality}.
\newblock {\em {IEEE} Transactions on Information Theory}, 66(1):147--154, jan
  2020.

\bibitem{sanz2010quantum}
M.~Sanz, D.~P{\'e}rez-Garc{\'i}a, M.~M. Wolf, and J.~I. Cirac.
\newblock {A quantum version of Wielandt's inequality}.
\newblock {\em IEEE Transactions on Information Theory}, 56(9):4668--4673,
  2010.

\bibitem{scarpa2018computational}
G.~Scarpa, A.~Molnar, Y.~Ge, J.~J. Garcia-Ripoll, N.~Schuch, D.~Perez-Garcia,
  and S.~Iblisdir.
\newblock {Projected {Entangled} {Pair} {States}: {Fundamental} {Analytical}
  and {Numerical} {Limitations}}.
\newblock {\em Physical Review Letters}, 125(21):210504, November 2020.

\bibitem{Schollw_ck_2011}
U.~Schollw{\"o}ck.
\newblock {The density-matrix renormalization group in the age of matrix
  product states}.
\newblock {\em Annals of Physics}, 326(1):96--192, jan 2011.

\bibitem{Schuch_2010}
N.~Schuch, J.~I. Cirac, and D.~P{\'e}rez-Garc{\'i}a.
\newblock {{PEPS} as ground states: Degeneracy and topology}.
\newblock {\em Annals of Physics}, 325(10):2153--2192, oct 2010.

\bibitem{PhysRevLett.98.140506}
N.~Schuch, M.~M. Wolf, F.~Verstraete, and J.~I. Cirac.
\newblock {Computational Complexity of Projected Entangled Pair States}.
\newblock {\em Phys. Rev. Lett.}, 98:140506, Apr 2007.

\bibitem{Shi_2006}
Y.-Y. Shi, L.-M. Duan, and G.~Vidal.
\newblock {Classical simulation of quantum many-body systems with a tree tensor
  network}.
\newblock {\em Physical Review A}, 74(2), aug 2006.

\bibitem{Str:RelativeBilComplMatMult}
V.~Strassen.
\newblock {Relative bilinear complexity and matrix multiplication}.
\newblock {\em J. Reine Angew. Math.}, 375/376:406--443, 1987.

\bibitem{Sylv:PrinciplesCalculusForms}
J.~J. Sylvester.
\newblock {On the principles of the calculus of forms}.
\newblock {\em Cambridge and Dublin Math. J.}, 7:52--97, 1852.

\bibitem{verstraete2004renormalization}
F.~Verstraete and J.~I. Cirac.
\newblock {Renormalization algorithms for quantum-many body systems in two and
  higher dimensions}.
\newblock {\em arXiv:cond-mat/0407066}, 2004.

\bibitem{Verstraete_2006}
F.~Verstraete, M.~M. Wolf, D.~Perez-Garcia, and J.~I. Cirac.
\newblock {Criticality, the Area Law, and the Computational Power of Projected
  Entangled Pair States}.
\newblock {\em Physical Review Letters}, 96(22), jun 2006.

\bibitem{Vidal_2007}
G.~Vidal.
\newblock {Entanglement Renormalization}.
\newblock {\em Physical Review Letters}, 99(22), nov 2007.

\bibitem{Vidal_2008}
G.~Vidal.
\newblock {Class of Quantum Many-Body States That Can Be Efficiently
  Simulated}.
\newblock {\em Physical Review Letters}, 101(11), sep 2008.

\bibitem{White_1992}
S.~R. White.
\newblock {Density matrix formulation for quantum renormalization groups}.
\newblock {\em Physical Review Letters}, 69(19):2863--2866, nov 1992.

\bibitem{xie_tensor_2014}
Z.~Y. Xie, J.~Chen, J.~F. Yu, X.~Kong, B.~Normand, and T.~Xiang.
\newblock {Tensor {Renormalization} of {Quantum} {Many}-{Body} {Systems}
  {Using} {Projected} {Entangled} {Simplex} {States}}.
\newblock {\em Physical Review X}, 4(1):011025, 2014.

\end{thebibliography}


\begin{thebibliography}{10}

\bibitem{BalBerChrGes:PartiallySymRkW}
E.~Ballico, A.~Bernardi, M.~Christandl, and F.~Gesmundo.
\newblock {On the partially symmetric rank of tensor products of W-states and
  other symmetric tensors}.
\newblock {\em Atti Accad. Naz. Lincei Rend. Lincei Mat. Appl.}, 30:93--124,
  2019.

\bibitem{Bini_1980}
D.~Bini, G.~Lotti, and F.~Romani.
\newblock {Approximate Solutions for the Bilinear Form Computational Problem}.
\newblock {\em {SIAM} Journal on Computing}, 9(4):692--697, nov 1980.

\bibitem{ChrGesJen:BorderRankNonMult}
M.~Christandl, F.~Gesmundo, and A.~K. Jensen.
\newblock {Border rank is not multiplicative under the tensor product}.
\newblock {\em SIAM J. Appl. Alg. Geom.}, 3:231--255, 2019.

\bibitem{Christandl2017_TR}
M.~Christandl, A.~K. Jensen, and J.~Zuiddam.
\newblock {Tensor rank is not multiplicative under the tensor product}.
\newblock {\em Linear Algebra and its Applications}, 543:125--139, 2018.

\bibitem{christandl2018tensor}
M.~Christandl, A.~Lucia, P.~Vrana, and A.~H. Werner.
\newblock {Tensor network representations from the geometry of entangled
  states}.
\newblock {\em SciPost Phys.}, 9(3):42, 2020.

\bibitem{efremenko94barriers}
K.~Efremenko, A.~Garg, R.~Oliveira, and A.~Wigderson.
\newblock {Barriers for Rank Methods in Arithmetic Complexity}.
\newblock In {\em {9th Innovations in Theoretical Computer Science Conference
  (ITCS 2018)}}. Schloss Dagstuhl-Leibniz-Zentrum fuer Informatik, 2018.

\bibitem{Hackbusch_2019}
W.~Hackbusch.
\newblock {A {Note} on {Nonclosed} {Tensor} {Formats}}.
\newblock {\em Vietnam Journal of Mathematics}, 48(4):621--631, December 2020.

\bibitem{Haferkamp_2020}
J.~Haferkamp, D.~Hangleiter, J.~Eisert, and M.~Gluza.
\newblock {Contracting projected entangled pair states is average-case hard}.
\newblock {\em Physical Review Research}, 2(1), jan 2020.

\bibitem{hoorfar2008inequalities}
A.~Hoorfar and M.~Hassani.
\newblock {Inequalities on the Lambert W function and hyperpower function}.
\newblock {\em J. Inequal. Pure and Appl. Math}, 9(2):5--9, 2008.

\bibitem{Lan:TensorBook}
J.~M. Landsberg.
\newblock {\em {Tensors: {G}eometry and {A}pplications}}, volume 128 of {\em
  {Graduate Studies in Mathematics}}.
\newblock American Mathematical Society, Providence, RI, 2012.

\bibitem{Landsberg_geometry}
J.~M. Landsberg, Y.~Qi, and K.~Ye.
\newblock {On the Geometry of Tensor Network States}.
\newblock {\em Quantum Info. Comput.}, 12(3--4):346--354, March 2012.

\bibitem{Michalek_2019_Wielandt}
M.~Michalek and Y.~Shitov.
\newblock {Quantum Version of Wielandt's Inequality Revisited}.
\newblock {\em {IEEE} Transactions on Information Theory}, 65(8):5239--5242,
  aug 2019.

\bibitem{Orus_practical}
R.~Or{\'u}s.
\newblock {A practical introduction to tensor networks: Matrix product states
  and projected entangled pair states}.
\newblock {\em Annals of Physics}, 349:117--158, oct 2014.

\bibitem{mapsgarcia}
D.~Perez-Garcia, F.~Verstraete, M.~M. Wolf, and J.~I. Cirac.
\newblock {Matrix Product State Representations}.
\newblock {\em Quantum Info. Comput.}, 7(5):401--430, July 2007.

\bibitem{Quarteroni_2006}
A.~Quarteroni, R.~Sacco, and F.~Saleri.
\newblock {\em {Numerical Mathematics}}.
\newblock Springer New York, 2007.

\bibitem{Rahaman_2020}
M.~Rahaman.
\newblock {A New Bound on Quantum Wielandt Inequality}.
\newblock {\em {IEEE} Transactions on Information Theory}, 66(1):147--154, jan
  2020.

\bibitem{sanz2010quantum}
M.~Sanz, D.~P{\'e}rez-Garc{\'i}a, M.~M. Wolf, and J.~I. Cirac.
\newblock {A quantum version of Wielandt's inequality}.
\newblock {\em IEEE Transactions on Information Theory}, 56(9):4668--4673,
  2010.

\bibitem{PhysRevLett.98.140506}
N.~Schuch, M.~M. Wolf, F.~Verstraete, and J.~I. Cirac.
\newblock {Computational Complexity of Projected Entangled Pair States}.
\newblock {\em Phys. Rev. Lett.}, 98:140506, Apr 2007.

\bibitem{Str:RelativeBilComplMatMult}
V.~Strassen.
\newblock {Relative bilinear complexity and matrix multiplication}.
\newblock {\em J. Reine Angew. Math.}, 375/376:406--443, 1987.

\bibitem{Sylv:PrinciplesCalculusForms}
J.~J. Sylvester.
\newblock {On the principles of the calculus of forms}.
\newblock {\em Cambridge and Dublin Math. J.}, 7:52--97, 1852.

\bibitem{Verstraete_2006}
F.~Verstraete, M.~M. Wolf, D.~Perez-Garcia, and J.~I. Cirac.
\newblock {Criticality, the Area Law, and the Computational Power of Projected
  Entangled Pair States}.
\newblock {\em Physical Review Letters}, 96(22), jun 2006.

\end{thebibliography}

\end{document}


\begin{center}
\textbf{\large Supplemental Material}
\end{center}
This is the supplemental material to the article \emph{Optimization at the boundary of the tensor network variety}. In Section~\ref{sec:repweight}, we discuss border rank and matrix product states representations of the weight states, which are used to define the ansatz class described in the main text. In Section~\ref{sec:efficiencyseparations}, we establish results on the separation between bond and border bond dimension of certain states, with and without imposing translational invariance.
In Section \ref{sec:approximaterep},
we discuss approximating states on the boundary of the tensor network variety by ones in the interior in more detail.
Finally, in Section \ref{sec:stableinterpolation}, we discuss how to perform stable polynomial interpolation for states in the new ansatz class.

\section{Representations for weight states}\label{sec:repweight}
This section discusses border rank and MPS representations for the weight states. For the convenience of the reader, we recall the definition of the (unnormalized) weight states from the main text:
\begin{defi}[Weight states]
Given $a,L,\mathfrak{d}\in\N_0$ the \emph{weight state} of weight $a$ on $L$ sites of local dimension $\frakd+1$ is:
\begin{align*}
     \ket{\chi_{a,\mathfrak{d},L}}=\sum\limits_{\substack{i_1+i_2+\ldots+i_L=a \\0\leq i_1,\ldots,i_L\leq \mathfrak{d}}}  \ket{i_1,i_2,\ldots,i_L} \in \left( \C^{\mathfrak{d}+1}\right)^{\otimes L} .
\end{align*}
Write $\ket{\chi_{a,L}}=\ket{\chi_{a,a,L}}$.
\end{defi}

A state $\ket{\psi}\in\lb \C^{d}\rb^{\otimes L}$ admits a representation of tensor rank $r$~\cite{Lan:TensorBook} if there exist product vectors $\ket{u_1},\ldots,\ket{u_r}\in \lb \C^{d}\rb^{\otimes L}$ such that
\begin{align}\label{equ:rankdecomposition}
\ket{\psi}=\ket{u_1} + \cdots + \ket{u_r}.
\end{align}

A state $\ket{\psi}\in\lb \C^{d}\rb^{\otimes L}$ admits a representation of border rank $r$~\cite{Bini_1980,christandl2018tensor} if there exist product vectors $\ket{u_1(\epsilon)},\ldots,\ket{u_r(\epsilon)}\in \lb \C^{d}\rb^{\otimes L}$, depending polynomially on $\eps$, and a nonnegative integer $a$ such that
\begin{align}\label{equ:borderrankdecomposition}
\ket{\psi}= \lim _{\epsilon\to0}\frac{1}{\epsilon^a} \lb \ket{u_1(\epsilon)} + \cdots + \ket{u_r(\epsilon)}\rb.
\end{align}

The minimal integer $r$ such that $\ket{\psi}$ has a representation of tensor rank (resp. border rank) is called tensor rank (resp. border rank) of $\ket{\psi}$, denoted $\rmR( \ket{\psi})$ (resp. $\uR(\ket{\psi})$).

Note that $\ket{\chi_{a,\mathfrak{d},L}}$ can be obtained from $\ket{\chi_{a,L}}$ applying local projection maps $P_{\mathfrak{d}}:\C^{a+1}\to \C^{\mathfrak{d}+1}$ defined by 
\begin{align*}
    P_{\mathfrak{d}}=\textsum_{k=0}^{\mathfrak{d}}\ketbra{k}{k}.
\end{align*}
In particular, a border rank $r$ representation of $\ket{\chi_{a,L}}$ immediately provides a border rank $r$ representation of $\ket{\chi_{a,\mathfrak{d},L}}$. Thus, we focus on efficient representations of $\ket{\chi_{a,L}}$.

\begin{prop}[Border rank of weight states]\label{thm:degwweight}
Let $\ket{\chi_{a,L}}\in\left( \C^{a+1}\right)^{\otimes L}$ be a weight state. For $\epsilon \in \C$, let $\ket{\phi(\epsilon)}\in\C^{a+1}$ be the state
\begin{align*}
    \ket{\phi_a(\epsilon)}=\sum\limits_{i=0}^a\epsilon^i\ket{i}.
\end{align*}
Then
\begin{equation}\label{eqn: border rank decomp weight}
    \ket{\chi_{a,L}}= \frac{1}{a!} \lim _{\epsilon \to 0}\frac{1}{\epsilon^a} \left[ \sum_{j=0}^a (-1)^{a-j}{\binom{a}{j}}\ket{\phi(j\epsilon)}^{\otimes L}\right].
\end{equation}
In particular, $\uR(\ket{\chi_{a,L}}) = a+1$.
\end{prop}
\begin{proof}
The lower bound is immediate via a standard flattening argument, see e.g.~\cite[Section 7.2]{BalBerChrGes:PartiallySymRkW} or~\cite{efremenko94barriers}.
We verify \eqref{eqn: border rank decomp weight} via a direct calculation:
\begin{align*}
\sum_{j=0}^a (-1)^{a-j}{\binom{a}{j}}\ket{\phi(j\epsilon)}^{\otimes L} &= \sum_{j=0}^a (-1)^{a-j} \binom{a}{j} \left[ \textsum_{i=0}^a (j \epsilon)^i \ket{i} \right]^{\otimes L}  \\
&=\sum_{j=0}^a (-1)^{a-j} \binom{a}{j} \left[ \textsum_{i_1 \vvirg i_L } j^{i_1 + \cdots +i_L} \eps^{i_1 + \cdots +i_L} \ket{i_1 \vvirg i_L} \right]  \\
&=\sum_{j=0}^a (-1)^{a-j} \binom{a}{j} \sum_{\alpha \geq 0} j^\alpha \eps^\alpha \left[ \textsum_{i_1 + \cdots + i_L =\alpha} \ket{i_1 \vvirg i_L} \right]  \\
&= \sum_{\alpha \geq 0} \eps^\alpha  \ket{ \chi_{\alpha,a,L}} \left[ \textsum_{j=0}^a  (-1)^{a-j} \binom{a}{j} j^\alpha \right].
\end{align*}
To conclude the proof we show that the coefficients $c_{\alpha,a} = \textsum_{j=0}^a  (-1)^{a-j} \binom{a}{j} j^\alpha$ satisfy $c_{\alpha,a} = 0$ if $\alpha < a$ and $c_{a,a} = a!$. To see this, we use a double induction on $a$ and $\alpha$.

If $(\alpha,a) = (0,0)$ then clearly $c_{\alpha,a}=1$, and if $\alpha = 0$ and $a > 0$, then $c_{\alpha,a} =  \textsum_{j=0}^a  (-1)^{a-j} \binom{a}{j} = (1-1)^a = 0$ by Newton's binomial formula.

Suppose $\alpha >0$; then the $0$-th term of the summation is $0$. We obtain
\begin{align*}
 c_{\alpha,a} & = \sum_{j=1}^a (-1)^{a-j} \binom{a}{j} j^\alpha  \\ 
 &=\sum_{j=1}^a (-1)^{a-j} \frac{a!}{j!(a-j)!} j^\alpha  \\
 &=a \sum_{j=1}^a (-1)^{(a-1)-(j-1)} \frac{(a-1)!}{(j-1)!(a-j)!} j^{\alpha-1}   \\
 &=a \sum_{j=0}^{a-1} (-1)^{(a-1)-j} \frac{(a-1)!}{j!((a-1)-j)!} (j+1)^{\alpha-1}   \\
 &=a \sum_{j=0}^{a-1} (-1)^{(a-1)-j} \frac{(a-1)!}{j!((a-1)-j)!} \left[ \sum_{\beta=0}^{\alpha-1} \binom{\alpha-1}{\beta} j^\beta  \right] =\\
 &=a \sum_{\beta=0}^{\alpha-1} \binom{\alpha-1}{\beta} \sum_{j=0}^{a-1} (-1)^{(a-1)-j} \binom{a-1}{j} j^\beta = a \sum_{\beta=0}^{\alpha-1} \binom{\alpha-1}{\beta}  c_{\beta, a-1}.
\end{align*}
By the inductive hypothesis, $c_{\beta,a-1} =0$ when $\beta < \alpha-1$ and $c_{a-1,a-1} = (a-1)!$. Since $\beta \leq \alpha-1$, we deduce that the summation above vanishes whenever $\alpha < a$. When $\alpha = a$, the summation reduces to $c_{\alpha, a} = a c_{\alpha-1,a-1} = a(a-1)! = a!$. This concludes the proof. 
\end{proof}

A consequence of Proposition \ref{thm:degwweight} is that $\uR(\ket{\chi_{a,\frakd,L}}) \leq a+1$, and in particular it is bounded from above independently of $L$. Moreover, all product states in the border rank expression of Proposition \ref{thm:degwweight} are symmetric. Regarding tensor rank, it is a classical fact \cite{Sylv:PrinciplesCalculusForms} that $\rmR( \ket{\chi_{1,L}}) = L+1$, which implies that $\rmR(\ket{\chi_{a,L}}) \geq L+1$ grows (at least) linearly in the system size $L$; on the other hand, since the border rank expression for $\ket{\chi_{a,L}}$ has degree $L-a$ in $\eps$, a standard interpolation argument (see, e.g., \cite[Thm. 8]{Christandl2017_TR} and \cite[Prop. 6.2]{ChrGesJen:BorderRankNonMult}) shows that $\rmR(\ket{\chi_{a,L}}) \leq (L-a+1) (a+1)$.

The following result provides a matrix product state representation with open boundary conditions for the weight states. As before, we only deal with the case where $\frakd=a$. This representation gives rise to the MPS contraction strategy described in the section \emph{Performing  computations  in  the  ansatz  class} of the main text.

\begin{lemma}\label{lemma: MPS for chi}
Fix $a,L\in\N$ and let $P_L$ be the graph consisting of an open chain with $L$ nodes indexed by $\{1 \vvirg L\}$. For $k = 1 \vvirg L$, define a local map $\mathcal{A}^{k}:\C^{a+1}\otimes \C^{a+1}\to \C^{a+1}$ with $\mathcal{A}^{k}=\sum\limits_{j=0}^{a}\ket{j}\otimes A^{(k)}_{j}$ where 
 \begin{align*}
 A^{(1)}_j &= \ket{j}\otimes \bra{j}, \\
 A^{(k)}_{j}&=\sum \limits_{i=0}^{a-j}\ket{i}\otimes \bra{i+j} \text{ for }  2\leq k\leq L-1 ,\\
 A^{(L)} _j &=\sum_{i=0}^{a} \ket{a-j}\otimes \bra{i}.
 \end{align*}
Let $\ket{\phi} = (\textbigotimes_{k =1}^L \calA_k ) \ket{\Omega^{(P_L)}_{a+1}}$ be the resulting tensor network state. Then $\ket{\phi} = \ket{\chi_{a,L}}$.
\end{lemma}
\begin{proof}
Write $A_j$ for $A_j^{(k)}$ when $k = 2 \vvirg L-1$ and define $A_j = 0$ for $j > a$. It is easy to see that  $A_jA_{j'}=A_{j+j'}$ for all $j,j'$. Therefore, if $i_2+i_3+\ldots+i_{L-1}>a$, we obtain
\begin{align*}
\langle i_1\ldots i_L| \phi\rangle=0.
\end{align*}
Therefore, suppose $i_2+i_3+\ldots+i_{L-1} = b \leq a$, so that $A_{i_2}A_{i_3}\ldots A_{i_{L-1}}=A_b$. Then
\begin{align*}
\langle i_1\ldots i_L| \phi\rangle=1
\end{align*}
if $i_1+i_{L}= a-b$ and $0$ otherwise. This concludes the proof. 
\end{proof}

Regarding matrix product states on the ring $C_L$, Lemma \ref{lemma: MPS for chi} implies that $\bTNS^{C_L}_{D,a,\frakd,d} \subseteq \TNS^{C_L}_{(a+1)D,d}$, relying simply on the fact that $\ket{\Omega^{(C_L)}_{D(a+1)}} =\ket{ \Omega^{(C_L)}_D} \otimes \ket{\Omega^{(P_L)}_{a+1}} \otimes \ket{\Omega^{(e)}_{a+1}}$, where $e$ is the edge which makes the chain $P_L$ into the ring $C_L$.

\section{Separations in efficiency of representations}\label{sec:efficiencyseparations}
As mentioned in the main text, for matrix product states representations on the cycle $C_L$, the maximal separation between the border bond dimension and bond dimension is at most quadratic. For the sake of completeness, we now provide a proof of this folklore result.

\begin{prop}[Separation between border bond and bond dimension]\label{prop: bond quadratic in border bond}
Let $\ket{\psi}\in\lb \C^d\rb^{\otimes L}$ be a state with $\bbond^{C_L} (\ket{\psi}) = D$. Then $\bond^{C_L}(\ket{\psi}) \leq D^2$. In particular
\[
 \cTNS^{C_L}_{D} \subseteq \TNS^{C_L}_{D^2}.
\]

\end{prop}
\begin{proof}
Let $P_L$ be the chain obtained from $C_L$ by removing a single edge $e_L$. Then $\TNS^{P_L}_{D^2}$ is a closed set \cite{Landsberg_geometry} and clearly $\TNS^{P_L}_{D^2} \subseteq \TNS_{D^2}^{C_L}$. We show that $\cTNS^{C_L}_D \subseteq \TNS^{P_L}_{D^2}$ and this will conclude the proof. 

Note that $\TNS^{P_L}_{D^2}$ can be interpreted as having two copies of maximally entangled stated of bond dimension $D$ on the chain $P_L$. By performing an entanglement swapping protocol, we can then generate a maximally entangled state between sites $1$ and $L$ with one of the maximally entangled states, which shows that we can generate any TNS on the cycle by starting with two copies on the chain.

More formally, let $ T : \bbC^D\otimes \bbC^D \to \bbC$ be the map defined by $T= \sum_1^D \bra{i} \otimes \bra{i}$. Define a family of linear maps $(\calA_k : k =1 \vvirg L)$ by $\calA^1 = \calA^k = \id_{\bbC^D}$ and $\calA^k = T$ for $k =2 \vvirg L-1$. Then $(\bigotimes_k \calA^k) \ket{\Omega^{(P_L)} _D }= \ket{\Omega^{(e_L)}_D}$. 

Since $\ket{\Omega^{(C_L)}_{D}} = \ket{\Omega^{(P_L)}_{D}} \otimes \ket{\Omega^{(e_L)}_D}$, one immediately has $\ket{\Omega^{(C_L)}_{D} }\in \TNS_{D^2}^{P_L}$ by applying the family $(\calA_k : k =1 \vvirg L)$ on one factor of $\ket{\Omega^{(P_L)}_{D^2}} = \ket{{\Omega^{(P_L)}_{D}}}^{\otimes 2}$. Therefore, 
\[
 \TNS_D^{C_L} \subseteq \TNS_{D^2}^{P_L}.
\]
Passing to the closures, we obtain $ \cTNS_D^{C_L} \subseteq \cTNS_{D^2}^{P_L}$, and since $\cTNS_{D^2}^{P_L} = \TNS_{D^2}^{P_L}$ is closed we conclude.
\end{proof}

The proof above is purely topological and not constructive: it only ensures the existence of a matrix product states representation of bond dimension $D^2$ under the hypothesis of the existence of a border bond dimension $D$ representation. 

The same argument can be generalized to arbitrary graphs as follows. Given a graph $G$, let $T$ be a spanning tree of $G$, i.e., a tree on all vertices of $G$. In particular, if $e$ is an edge of $G$ which is not an edge of $T$, then $T \cup \{e \}$ contains a cycle; the entanglement swapping argument of Proposition \ref{prop: bond quadratic in border bond} shows that $\Omega^{(e)}_D \in \TNS_D^{T}$. In this way, one can obtain the maximally entangled states $\Omega_D^{(e)}$ corresponding to all edges of $G$ which are not edges of $T$.  However, without any assumptions on $G$, one might need a full $\Omega_D^{(T)}$ to generate each edge.

Since $T$ does not contain any cycle, the set $\TNS_D^T$ is a closed set and we conclude
\[
\cTNS^{G}_D \subseteq \TNS^T_{D^\eta} \subseteq \TNS^G_{D^\eta} 
\]
where $\eta$ is the number of edges of $G$ which are not edges of $T$. As any tree with $|V|$ vertices has $|E|=|V|-1$ edges, we conclude that  for a graph $G=(V,E)$
\begin{align*}
\cTNS^{G}_D \subseteq \TNS^G_{D^{|E|-|V|+2}}. 
\end{align*}

\subsection{Examples of system-size dependent separations in efficiency of representation}\label{sec:analyticalexamples}
In this section, we discuss an example of a state $\ket{\psi_L} \in (\bbC^3)^{\otimes L}$ for odd $L$ which admits a border bond dimension $2$ translation invariant matrix product state representation with approximation degree $a=1$, yet having translation invariant bond dimension growing with $L$. More precisely, 
\begin{align*}
\bbond^{\text{TI-}C_L}(\ket{\psi_L}) &= 2 \text{ (with $a = \frakd = 1$)} \\
\bond^{\text{TI-}C_L}(\ket{\psi_L}) & = \Omega \left( \textfrac{L^{1/3}}{\log(L)} \right) .
\end{align*}
Moreover, the state $\ket{\psi_L}$ arises as ground state of a translation invariant local (nearest neighbor) Hamiltonian on the cycle. In particular, we obtain $\ket{\psi_L} \in \bTNS^{\text{TI-}C_L}_{2,1,1,3}$ and $\ket{\psi_L} \notin \TNS^{\text{TI-}C_L}_{D,3}$ if $D$ grows slower than $\Omega( \frac{L^{1/3}}{\log(L)})$.

In the non-translation invariant case, there are no known examples of separations that are larger than a constant factor. 

The separation in the translation invariant setting, and in particular the lower bound on $\bond^{\text{TI-}C_L}(\ket{\psi})$ relies on the analogous result for the $W$-state, proved with the same argument as \cite[Corollary 1]{mapsgarcia} via the results of \cite{sanz2010quantum,Rahaman_2020,Michalek_2019_Wielandt}.

\begin{lemma}\label{lem:repwstate}
Let $C_L$ the cycle on $L$ nodes. Then 
\begin{align*}
 (a+1)L\geq\bond^{\text{TI-}C_L}(\ket{\chi_{a,L}}) &= \Omega  \left(\frac{L^{1/3}}{\log(L)}\right) ,\\
\bbond^{\text{TI-}C_L} (\ket{\chi_{a,L}}) &\leq a+1.
\end{align*}
\end{lemma}
\begin{proof}
The upper bound $\bbond^{\text{TI-}C_L} (\ket{\chi_{a,L}})\leq a+1$ is obtained directly from the border rank expression of Proposition \ref{thm:degwweight}.  This is because the superposition of $a+1$ symmetric product states can be written as a TI-MPS of bond dimension $a+1$.

The upper bound $\bond^{\text{TI-}C_L}(\ket{\chi_{a,L}}) \leq (a+1)L$ is obtained symmetrizing the non-translation invariant representation obtained in Lemma \ref{lemma: MPS for chi}.

As for the lower bound $\bond^{\text{TI-}C_L}(\ket{\chi_{a,L}}) = \Omega  \left(\frac{L^{1/3}}{\log(L)}\right)$, first note that 
the (local and translation invariant) projection map $P^{\otimes L}$ where $P = \ketbra{0}{0} + \ketbra{1}{a}$ sends $\ket{\chi_{a,L}}$ to the $W$-state $\ket{\chi_{1,L}}$. In particular, lower bounds on $\bond^{\text{TI-}C_L}(\ket{\chi_{1,L}})$ are lower bounds for $\ket{\chi_{a,L}}$ as well.

In~\cite[Corollary 1]{mapsgarcia}, the authors related $\bond^{\text{TI-}C_L}(\ket{\chi_{1,L}})$ to quantum Wielandt inequalities~\cite{sanz2010quantum,Rahaman_2020,Michalek_2019_Wielandt}. Although the version of the inequality stated in~\cite[Conjecture 2]{mapsgarcia}, on which \cite[Corollary 1]{mapsgarcia} relies on, is only conjectured, in~\cite{Michalek_2019_Wielandt} the authors show that~\cite[Conjecture 2]{mapsgarcia} holds with $f(D)=\mathcal{O}(D^2\log(D))$.

Reproducing the proof of~\cite[Corollary 1]{mapsgarcia} provides 
\begin{align*}
    (D)^3\log(D)^3=\Omega(L),
\end{align*}
whenever $D \geq \bond^{\text{TI-}C_L}(\ket{\chi_{1,L}})$. This implies $ \bond^{\text{TI-}C_L}(\ket{\chi_{1,L}}) =\Omega(e^{\mathcal{W}(L^{1/3})})$, where $\mathcal{W}$ is the Lambert $\mathcal{W}$ (see, e.g., \cite{hoorfar2008inequalities}). Indeed, one can show \cite[(1.1)]{hoorfar2008inequalities} that
\begin{align*}
\mathcal{W}(x)\geq \log(x)-\log \log(x)
\end{align*}
for all $x\geq e$, which implies that $e^{\mathcal{W}(L^{1/3})}=\Omega\left(\frac{L^{1/3}}{\log(L)}\right) $.
\end{proof}

From Lemma \ref{lem:repwstate}, already $\ket{\chi_{a,L}}$ provides an example of a state for which we have a separation. 

However, in the main text we presented a local translationally invariant Hamiltonian whose unique translationally invariant ground state presents this behaviour as well. Let us recall its definition and discuss some of its properties. The weight states themselves give separations in the required bond dimension and will be used as stepping stones to prove separations for these examples. But the next examples require a bond dimension $D>1$ to be represented in the ansatz class introduced  class, unlike the weight states.

Let $P :\bbC^3 \otimes \bbC^3 \to \bbC^3 \otimes \bbC^3$ be the projection defined by
\begin{align}\label{equ:definitonprojection}
P=\ketbra{01}{01}+\ketbra{10}{10}+\ketbra{02}{02}+\ketbra{21}{21};
\end{align}
denote by $P_{i,i+1}$ the linear map $P$ acting on the two factors of $(\bbC^3)^{\otimes L}$ corresponding to sites $i$ and $i+1$ of the ring $C_L$. 
Define $H_{i,i+1} = \id - P_{i,i+1} + \ketbra{2}{2}_i$, where the third summand acts on the $i$-th site. Let $H = \sum_i H_i$: this is a diagonal, translationally invariant Hamiltonian. 

We determine its translationally invariant ground state in the case where $L$ is odd.
\begin{prop}
Let $L$ be odd. Then $H$ has a unique translationally invariant ground state given by
\begin{align}\label{equ:TIgroundstate}
\ket{\psi_L}=\frac{1}{\sqrt{L}}\sum\limits_{i=0}^{L-1}S^i\ket{2101010\ldots 10},
\end{align}
where $S$ is the cyclic permutation of the tensor factors, i.e., the shift operator.
\end{prop}
\begin{proof}
Since $H$ is diagonal, every product state in the standard basis is and eigenstate for $H$. Moreover, using the fact that the $P_{i,i+1}$ commute, one obtains that the spectrum of $H$ is contained in the set
\begin{align*}
\left\{\ell+\frac{m}{2L}|\ell,m \in\N_{0}\right\}.
\end{align*}
Given a product state in the standard basis, we compute its energy,i.e. the corresponding eigenvalue of $H$. 

Note that every state involving two adjacent $\ket{0}$'s or two adjacent $\ket{1}$'s has energy at least $1$, since $1$ is the eigenvalue of the corresponding $H_i$. Moreover, since $L$ is odd, if the state does not involve any $\ket{2}$, there are necessarily adjacent $\ket{0}$'s or $\ket{1}$'s. Every state involving at least two $\ket{2}$'s has energy at least $2 \cdot 1/(2L)$, arising from the $\ketbra{2}{2}$ summands in $H$. There are only two basis state, up to cyclic permutation, involving exactly one $\ket{2}$ and having no adjacent $\ket{0}$'s nor $\ket{1}$'s: they are $\ket{20101 \ldots 01}$ and $\ket {21010 \ldots 10}$. The first has energy at least $1$ arising from $H_{12}$. The state $\ket {21010 \ldots 10}$ has energy $1/(2L)$. 

We conclude that $\{ S^k\ket{2101010\ldots 10} : k = 0 \vvirg L-1\}$ span the ground state space of $H$. The only translation invariant state in this span is $\ket{\psi_L}$.
\end{proof}

The following result gives upper and lower bounds for the translation invariant bond and border bond dimension of $\ket{\psi_L}$:
\begin{prop}
Let $\ket{\psi_L} \in (\bbC^3)^{\otimes L}$ be the state of \eqref{equ:TIgroundstate}. Then
 \begin{align*}
\bbond^{\text{TI-}C_L}(\ket{\psi_L}) &= 2;\\
  \bond^{\text{TI-}C_L}(\ket{\psi_L}) &= \Omega\lb \frac{L^{\frac{1}{3}}}{\log(L)}\rb.
 \end{align*}
In fact, $\ket{\psi_L} \in \bTNS^{\text{TI-}C_L}_{2,1,1,3}$ and, for $L$ large enough, $\ket{\psi_L} \notin \TNS^{\text{TI-}C_L}_{D,3}$ if $D$ does not depend on $L$.
\end{prop}
\begin{proof}
Let $\calQ: \bbC^3 \to \bbC^2$ be the local map defined by 
\begin{align*}
\calQ=\ketbra{0}{1}+\ketbra{0}{0}+\ketbra{1}{2}.
\end{align*}
Then $\calQ^{\otimes L} \ket{\psi_L}=\ket{\chi_{1,L}}$ is the $W$-state. In particular, the lower bound for $\ket{\chi_{1,L}}$ obtained in Lemma~\ref{lem:repwstate} holds for $\bond^{\text{TI-}C_L}(\ket{\psi_L})$ as well.

Now, define $\mathcal{A}(\eps): \C^{2}\otimes \C^{ 2} \to\C^3$ by
\begin{align*}
\mathcal{A}(\eps)=\ketbra{0}{01}+\ketbra{1}{10}+\eps \ketbra{2}{00}.
\end{align*}
One can readily check that 
\begin{align*}
\lim_{\eps \to 0} \eps^{-1} \lb \mathcal{A}(\eps)^{\otimes L}\rb \ket{\Omega_2^{(C_L)}}=\ket{\psi_L},
\end{align*}
giving a representation of $\ket{\psi_L}$ of border bond dimension $2$ and approximation degree $1$. In particular, $\ket{\psi_L} \in \bTNS^{C_L}_{2,1,1,3}$ and concludes the proof.
\end{proof}
One can define explicitly a local map $\calB : (\C^{2}\otimes \C^{ 2} )  \otimes \bbC^2 \to \bbC^3$ by 
\[
(\ketbra{0}{01}+\ketbra{1}{10}) \otimes \bra{0}+\ketbra{2}{00} \otimes \bra{1}
\]
and observe $\ket{\psi_L} = \calB^{\otimes L}\ket{ \Omega^{(C_L)}_2} \otimes \ket{\chi_{1,L}}$.

This shows that there are unbounded separations in the bond dimension required to represent states that arise as ground states of local Hamiltonians in the new ansatz class for translationally invariant matrix product state when compared with standard matrix product states. 

\subsection{An explicit example of separation between bond and border bond dimension}

For small system size and small local dimension, several explicit examples of separation between bond and border bond dimension are known. In \cite{Str:RelativeBilComplMatMult}, Strassen proved that the GHZ state of level three on three parties, $\ket{\ghz_3} =3^{-\frac{1}{2}}\lb \ket{000} + \ket{111} + \ket{222}\rb$, satisfies $\bbond^{C_3} (\ket{\ghz}) = 2$, realized by a degeneration with approximation degree $a=2$; in fact, one can achieve the same result with a degeneration of approximation degree $a=1$ and in addition in \cite{christandl2018tensor} it was proved that $\bond^{C_3} (\ket{\ghz}) = 3$, providing a separation. 

In \cite{christandl2018tensor}, a separation was shown for the state $\ket{\lambda} = \sum_{\sigma \in \frakS_3} (-1)^\sigma \ket{\sigma(0)\sigma(1)\sigma(2) } + |222\rangle$ as well; here $\frakS_3$ denotes the symmetric group on three elements and $(-1)^\sigma$ is the sign of a permutation $\sigma$; in particular $\sum_{\sigma \in \frakS_3} (-1)^\sigma \ket{\sigma(0)\sigma(1)\sigma(2) }$ is the unique (up to scaling) alternating tensor in $\bbC^3 \otimes \bbC^3 \otimes \bbC^3 $. It was proved that $\bbond^{C_3} (\ket{\lambda}) =2$ and $\bond^{C_3} = 3$. 

For these examples, comparing the numerical results in the standard tensor network ansatz class with the ones in the new ansatz class, we observe improvements both in the distance and in the convergence. However, in this case, the class $\bTNS^{C_3} _{2,1,1,3}$ is dense in $\bbC^3 \otimes \bbC^3 \otimes \bbC^3$, so the improvements occur for a trivial reason, as the optimization procedure can just ``walk straight'' to the global minimum. 

In order to see an example where the optimization procedure is not trivial, we consider the state 
\begin{equation}\label{eqn: tensor degen 3x3 mamu}
\begin{aligned}
 \ket{T} &=\sqrt{17}^{-1}( \ket{005}+\ket{016}+\ket{040}+\ket{126}+\ket{160}+\ket{227}+\ket{251}+\ket{262}+\ket{338}\\
 &+\ket{373}+\ket{384}+\ket{430}+\ket{501}+\ket{632}+\ket{703}+\ket{714}+\ket{824}) \in \bbC^9 \otimes \bbC^9 \otimes \bbC^9,
 \end{aligned}
\end{equation}
introduced in the \emph{Numerical examples} section of the main text. Lemma \ref{lemma: bond and border bond of T} below shows $\ket{T} \notin \TNS^{C_3}_{3,9}$ and $\ket{T} \in \bTNS^{C_3}_{3,1,9}$. Moreover, the set $\bar{\bTNS}^{C_3}_{3,1,9}$ is a proper subvariety of $\bbC^9 \otimes \bbC^9 \otimes \bbC^9$, therefore the new ansatz class is non-trivial. 

\begin{lemma}\label{lemma: bond and border bond of T}
Let $\ket{T}$ be the state defined in \eqref{eqn: tensor degen 3x3 mamu}. Then 
\begin{align*}
\bond^{C_3}{(\ket{T})} &\geq 4 , \\
\bbond^{C_3}{(\ket{T})}& = 3 \text{ and the degeneration has approximation degree $a = 1$}.
\end{align*}
In particular, $\ket{T} \notin \TNS^{C_3}_{3,9}$ and $\ket{T} \in \bTNS^{C_3}_{3,1,9}$.
\end{lemma}
\begin{proof}
Assume by contradiction $\bond^{C_3}(\ket{T}) \leq 3$ and observe that $\ket{\Omega^{(C_3)}_3} \in \bbC^9 \otimes \bbC^9 \otimes \bbC^9$, where each $\bbC^9$ has a local structure $\bbC^3 \otimes \bbC^3$. For $i=1,2,3$, let $\calA_i : (\bbC^3 \otimes \bbC^3) \to \bbC^9$ be the linear maps such that $(\bigotimes \calA^i) (\ket{\Omega^{(C_3)}_3}) = \ket{T}$. A direct calculation shows that there is no proper subspace $U \subsetneq \bbC^9$ such that $\ket{T} \in U \otimes \bbC^9 \otimes \bbC^9$ or $\ket{T} \in \bbC^9 \otimes U \otimes \bbC^9$ or $\ket{T} \in \bbC^9 \otimes \bbC^9 \otimes U$. Thus, the maps $\calA^i$ are invertible, and $\ket{T}$ is a point of the orbit of $\ket{\Omega^{(C_3)}_3}$ under the action of $GL_9 \times GL_9 \times GL_9$. As a consequence, the stabilizer of $\ket{T}$ and the one of $\ket{\Omega^{(C_3)}_3}$ are conjugate and in particular they have the same dimension. But this is not the case: a direct calculation shows the stabilizer of $\ket{\Omega^{(C_3)}_3}$ is $26$-dimensional whether the stabilizer of $\ket{T}$ is $28$-dimensional. This shows the lower bound $\bond^{C_3} (\ket{T}) \geq 4$. 

We provide a degeneration of $\ket{\Omega^{(C_3)}_3}$ with approximation degree $a=1$ which realizes $\ket{T}$. In other words, we determine three linear maps $\calA^{i}(\eps) : \bbC^3 \otimes \bbC^3 \to \bbC^9$ such that $\lim _{\eps \to 0} ( \frac{1}{\eps} \bigotimes _i \calA^i(\eps)) \ket{\Omega^{(C_3)}_3} = \ket{T}$. Write $\calA^i(\eps) = \calA_0^{i} +  \eps \calA_1^i$ and express $\calA_0^i$ and $\calA_1^i$ as matrices with entries in $\bbC^9$:
\begin{align*}
 \calA_0^1 = \calA_0^2 = \left(\begin{array}{ccc}
                          \ket{0} & \ket{1} & \\
                          & \ket{2} & \\
                          & & \ket{3}
                         \end{array} \right)
                         & \quad 
 \calA_0^3 = \left(\begin{array}{ccc}
                         & \ket{1}   & \ket{3} \\
                          & & \ket{4} \\
                          \ket{0} & \ket{2} & 
                         \end{array} \right) \\
 \calA_1^1 = \calA_1^2 = \left(\begin{array}{ccc}
                         & & \ket{4} \\
                         \ket{5} & & \ket{6} \\
                          \ket{7} & \ket{8} & 
                         \end{array} \right)
&          \quad               
\calA_1^3 = \left(\begin{array}{ccc}
                          \ket{5} &  & \\
                          \ket{6} & \ket{7} & \\
                          & & \ket{8}
                         \end{array} \right)
                         & \quad 
 \end{align*}
A direct calculation shows that $\lim _{\eps \to 0} ( \frac{1}{\eps} \bigotimes _i \calA^i(\eps)) \ket{\Omega^{(C_3)}_3} = \ket{T}$.
\end{proof}

\section{Approximate representations}\label{sec:approximaterep}
As we already showed, there exist states that arise as ground states of local Hamiltonians and admit a more efficient \emph{exact} representation in $\textrm{bTNS}$. However, by definition, if a state lies on the boundary of the TNS variety, it can be approximated arbitrarily well by ``usual" TNS, albeit with local maps whose entries diverge as we increase the precision. 

Most rigorous results available in the literature only ascertain that tensor network states of a polynomial bond dimension \emph{approximate} ground states well~\cite{Verstraete_2006}. Thus, it is of high importance to investigate the interplay between the speed at which the local maps diverge and how well we can approximate the target state by the restrictions arising from degenerations. This issue that was also raised in~\cite{Hackbusch_2019} recently. In other words, it is important to understand how small one should pick $\epsilon$ so that $\ket{\phi(\epsilon)}\simeq \ket{\phi(0)}$ for the task at hand. First, recall that to estimate most quantities of physical interest up to an additive error $\delta>0$,  such as the energy or two-point correlation functions of the ground state, it usually suffices to estimate states to a precision of $\mathcal{O}(\delta L^{-\alpha})$ in the Euclidean norm, where again $L$ is the system size and $\alpha\in\N_0$. This is because these quantities are expectation values of observables with operator norm  $\cO(L^{\alpha})$.

With that in mind, we are ready to discuss how to pick $\epsilon$ for a degeneration to ensure the desired precision. To gain some intuition about how the local maps behave in the regime of interest, we consider the case of the $W$-state, before considering the general setting.

Consider the family $\ket{\psi(\epsilon)}$ of TI-MPS on a cycle of size $L$ with bond dimension $2$ and matrices given by:
\begin{align}\label{equ:MPSdefining}
    A_0(\epsilon)= \left(\begin{matrix}1&0\\0& e^{\frac{\pi i}{L}}\end{matrix}\right), \quad A_1(\epsilon)=\left(\begin{matrix}\epsilon&0\\0&0\end{matrix}\right).
\end{align}
One can readily check that
\begin{align*}
    \lim\limits_{\epsilon\to 0}\epsilon^{-1}\ket{\psi(\epsilon)}=\ket{W},
\end{align*}
where $\ket{W}$ is the unnormalized $W$-state:
\begin{align*}
\ket{W}=\ket{100\ldots 0}+\ket{010\ldots 0}+\ldots+\ket{000\ldots 1}.
\end{align*}
Moreover, the transfer matrix $\mathcal{E}(\epsilon)$ of this family of MPS is given by:
\begin{align*}
   \mathcal{E}(\epsilon)= \left(\begin{matrix}1+\epsilon^2&0&0&0\\0& e^{\frac{\pi i}{L}}&0&0\\0&0& e^{\frac{\pi i}{L}}&0\\0&0&0&1\end{matrix}\right).
\end{align*}
Thus, we have that:
\begin{align*}
    \|\psi(\epsilon)\|^2=\tr{\mathcal{E}(\epsilon)^L}=(1+\epsilon^2)^L-1.
\end{align*}
It is also straightforward to compute the overlap of $\ket{\psi(\epsilon)}$ with the $W$-state, as:
\begin{align*}
    \scalar{W}{\psi(\epsilon)}=L\tr{A_0^{L-1}A_1}=L\epsilon.
\end{align*}
Combining these expressions and normalizing all involved states we have:
\begin{align}\label{equ:overlapWdeg}
    \left\|\frac{\ket{\psi(\epsilon)}}{\|\ket{\psi(\epsilon)}\|}-\frac{\ket{W}}{\sqrt{L}}\right\|^2=2\lb1-\frac{L\epsilon}{\lb L\lb \lb 1+\epsilon^2\rb^L-1\rb\rb^{\frac{1}{2}}}\rb=2\frac{\lb \lb 1+\epsilon^2\rb^L-1\rb^{\frac{1}{2}}-\sqrt{L}\epsilon}{\lb \lb 1+\epsilon^2\rb^L-1\rb^{\frac{1}{2}}}
\end{align}
Performing a Taylor expansion, we see that
\begin{align*}
2\frac{\lb \lb 1+\epsilon^2\rb^L-1\rb^{\frac{1}{2}}-\sqrt{L}\epsilon}{\lb \lb 1+\epsilon^2\rb^L-1\rb^{\frac{1}{2}}}=\cO(L\epsilon^2).
\end{align*}
Thus, in order to ensure that
\begin{align*}
 \left\|\frac{\ket{\psi(\epsilon)}}{\|\ket{\psi(\epsilon)}\|}-\frac{\ket{W}}{\sqrt{L}}\right\|^2\leq \frac{\delta^2}{L^{2\alpha}}
\end{align*}
for some $\alpha\in \N_0$, we need to pick $\epsilon=\cO(\delta L^{-\frac{2\alpha+1}{2}})$.

Inserting this bound in the definitions of the matrices in Eq. \eqref{equ:MPSdefining} and normalizing the underlying MPS, we see that in order to obtain an approximation up to an error $\delta L^{-\alpha}$ to the W-state in Euclidean norm we need to pick the matrices with entries scaling like:
\begin{align*}
        A_0= \lb\frac{e^{\log(L)\frac{2\alpha+1}{2}}}{\delta}\rb^{\frac{1}{L}}\left(\begin{matrix}1&0\\0& e^{\frac{\pi i}{L}}\end{matrix}\right), \quad A_1=\lb\frac{e^{-\log(L)\frac{2\alpha+1}{2}}}{\delta}\rb\left(\begin{matrix}1&0\\0&0\end{matrix}\right).
\end{align*}
Note that $A_0$ has entries of constant order, while $A_1$ has entries of inverse polynomial order for $\alpha$ fixed. Moreover, the norm of the underlying MPS will be of order $\delta L^{-\alpha}$, which implies that trying to approximate the $W$-state with this family of MPS leads to ill-conditioned MPS. Setting the vanishing entries to $0$ leads to a state that is \emph{orthogonal} to the state of interest.
We expect that this behavior is universal, i.e., that \emph{any} approximation of states on the boundary of the set of TNS of a given bond dimension up to inverse polynomial in system size  precision leads to tensors with entries that are polynomially small and ill-conditioned. 
Unfortunately, we still lack the tools to prove such statements for any approximation of a degeneration and can only prove it for some given degeneration, as we did for the $W$ state:
\begin{lemma}[Approximate representation from degeneration]\label{lem:approximate_rep}
Let $\mathcal{A}^k(\epsilon):\lb \C^{D}\rb^{\otimes k_v}\to \C^d$ be a family of local maps whose entries depend polynomially on $\epsilon$ defining a degeneration to $\ket{\phi_0}$ with $\|\phi_0\|=1$. Denote the expansion of 
\begin{align*}
\ket{\psi(\epsilon)}=\lb\bigotimes_{v\in V}\mathcal{A}^v(\epsilon)\rb\lb\bigotimes\limits_{y\in E} \ket{\Omega^y}   \rb
\end{align*}
as a polynomial in terms of $\epsilon$ by:
\begin{align}\label{equ:expansionhigherorder}
\ket{\psi(\epsilon)}=\epsilon^{a}\ket{\phi_0}+\sum\limits_{l=1}^{e}\epsilon^{a+l}\ket{\phi_l}
\end{align}
for some $a\in\N_0$ and set
\begin{align*}
\tau(\epsilon)=\sum\limits_{l=1}^{e}\epsilon^{l}\|\ket{\phi_l}\|,
\end{align*}
Then, for $\tau(\epsilon)\leq\frac{1}{2}$ we have:
\begin{align}\label{equ:firstapprox}
\left\|\frac{\ket{\psi(\epsilon)}}{\|\ket{\psi(\epsilon)}\|}-\ket{\phi_0}\right\|\leq 4\tau(\epsilon)
\end{align}
\end{lemma}
\begin{proof}

By the triangle inequality we have:
\begin{align}\label{equ:unnormalized2}
\left\|\frac{\ket{\psi(\epsilon)}}{\|\psi(\epsilon)\|}-\ket{\phi_0}\right\|\leq\left\|\frac{\ket{\psi(\epsilon)}}{\epsilon^a}-\ket{\phi_0}\right\|+\left\|\frac{\ket{\psi(\epsilon)}}{\|\ket{\psi(\epsilon)}\|}-\frac{\ket{\psi(\epsilon)}}{\epsilon^a}\right\|
\end{align}
Let us analyze the first term in the expression above. Expanding the polynomials we see that:
\begin{align}\label{equ:firsttriangle}
\left\|\frac{\ket{\psi(\epsilon)}}{\epsilon^a}-\ket{\phi_0}\right\|=\left\|\sum\limits_{l=1}^{e}\epsilon^{l}\ket{\phi_l}\right\|\leq \tau(\epsilon)
\end{align}
by a triangle inequality.
Let us now analyze the second term. We have:
\begin{align}\label{equ:normalized}
&\left\|\frac{\ket{\psi(\epsilon)}}{\|\ket{\psi(\epsilon)}\|}-\frac{\ket{\psi(\epsilon)}}{\epsilon^a}\right\|=\left|\frac{1}{\|\ket{\psi(\epsilon)}\|}-\frac{1}{\epsilon^a}\right|\|\ket{\psi(\epsilon)}\|=\left|\frac{1}{\epsilon^{-a}\|\ket{\psi(\epsilon)}\|}-1\right|\epsilon^{-a}\|\ket{\psi(\epsilon)}\|
\end{align}
Again by a triangle inequality and the assumption that $\|\phi_0\|=1$, it follows from eq.~\eqref{equ:expansionhigherorder} that:
\begin{align*}
\|\epsilon^{-a}\psi(\epsilon)\|\leq 1+\tau(\epsilon).
\end{align*}

Moreover, by a reverse triangle inequality
\begin{align*}
\left|\epsilon^{-a}\|\ket{\psi(\epsilon)}\|-\|\ket{\phi_0}\|\right|=\left|\epsilon^{-a}\|\ket{\psi(\epsilon)}\|-1\right|\leq\|\epsilon^{-a}\ket{\psi(\epsilon)}-\ket{\phi_0}\|\leq \tau(\epsilon)
\end{align*}
and we assumed that $\|\phi_0\|=1$. 
This yields
\begin{align*}
\epsilon^{-a}\|\psi(\epsilon)\|\geq 1-\tau(\epsilon)
\end{align*}
and we obtain
\begin{align*}
\left|\frac{1}{\epsilon^{-a}\|\ket{\psi(\epsilon)}\|}-1\right|\leq\frac{\tau(\epsilon)}{1-\tau(\epsilon)}.
\end{align*}
Together with Eq.~\eqref{equ:normalized} we conclude that
\begin{align}\label{equ:boundonnormalized}
\left\|\frac{\ket{\psi(\epsilon)}}{\|\ket{\psi(\epsilon)}\|}-\frac{\ket{\psi(\epsilon)}}{\epsilon^a}\right\|\leq \frac{\tau(\epsilon)(1+\tau(\epsilon))}{1-\tau(\epsilon)}\leq 3\tau(\epsilon),
\end{align}
as we assumed that $\tau(\epsilon)\leq 1/2$.
Inserting the bounds of Equation~\eqref{equ:boundonnormalized} and~\eqref{equ:firsttriangle} into~\eqref{equ:unnormalized2} we obtain the claim in~\eqref{equ:firstapprox}.

\end{proof}
Thus, we see that in order to obtain an approximation up to an error $\cO(\delta L^{-\alpha})$, it suffices to pick $\epsilon_0=\frac{\delta}{L^{\alpha}M}$, where $M=\max \|\phi_l\|$. Indeed, for this choice of parameters, we have $\tau(\epsilon_0)=\cO(\delta L^{-\alpha})$. Moreover, we have:
\begin{align*}
\|\ket{\psi(\epsilon_0)}\|=\cO\lb\frac{\delta^a}{L^{a\alpha}M^a}\rb.
\end{align*}
and we expect that the condition number of the underlying maps scales in a similar way. This implies, for instance, that in order to obtain an estimate of the energy of the normalized state $\ket{\psi(\epsilon_0)}/\|\psi(\epsilon_0)\|$, it will be necessary to perform the multiplication $\scalar{\psi(\epsilon_0)}{H\psi(\epsilon_0)}\times\|\psi(\epsilon_0)\|^{-2}$, which entails multiplying two numbers of order $L^{-2a}\delta^{2a}$ and $L^{2a}\delta^{-2a}$, respectively.
For $a$ constant this requires a polynomial precision in the computation of expectation values and in case $a$ is at least linear in system size, this requires exponential precision.

We conclude that approximating states arising from degenerations by restrictions requires a great amount of precision when computing expectation values and will lead to numerical instabilities as the system size or degree of the degenerations increases. On the other hand, our methods remain stable.

This is even more pronounced in the case of tensor networks that are hard to contract exactly. In those cases, it is not feasible to contract the restrictions to the precision we just established are necessary. However, as highlighted in Section~\ref{sec:stableinterpolation}, approximate contractions yield good approximations in bTNS.

However, note that in principle some of the low order terms of $\tau(\epsilon)$ may vanish as well, which would lead to better convergence estimates than the ones discussed above. Let us illustrate this phenomenon in the case of the weight states:

\subsection{Approximate representations for the weight states}

Let us analyze more closely the trade-off between the rank of border rank representations of weight states and the speed at which they converge to the state. As noted in~\cite{Hackbusch_2019}, the representation given in~\eqref{equ:MPSdefining} for the W-state is suboptimal w.r.t. to the convergence speed. Indeed, one can check that:
\begin{align}\label{equ:betterrepW}
\frac{1}{2\epsilon}\lb\lb \ket{0}+\epsilon \ket{1}\rb^{\otimes L}-\lb \ket{0}-\epsilon \ket{1}\rb^{\otimes L}\rb=\ket{W}+\cO(\epsilon^2),
\end{align}
while the degeneration given in~\eqref{equ:MPSdefining} only converges up to $\cO(\epsilon)$.
This improved representation is closely connected with the fact that we can also obtain the weight states as:
\begin{align}\label{equ:representationderivative}
\ket{\chi_{a,L}}=a!\frac{d^a}{d \epsilon^a}\ket{\Gamma_a(\epsilon)}\big |_{\epsilon=0},
\end{align}
where
\begin{align*}
\ket{\Gamma_a(\epsilon)}=(\ket{0}+\epsilon \ket{1}+\ldots+\epsilon^{a}\ket{a})^{\otimes L}.
\end{align*}
For instance, the degeneration given in Eq.~\eqref{equ:betterrepW} arises from the symmetric difference formula for the derivative of a function:
\begin{align*}
\frac{f(x+\epsilon)-f(x-\epsilon)}{2\epsilon}=f'(x)+\cO(\epsilon^2).
\end{align*}
Evaluating this formula for $\ket{\Gamma_a(\epsilon)}$ we obtain this improved degeneration. More generally, so-called central finite difference formulas~\cite[Section 10.10]{Quarteroni_2006} allow us to estimate the $a$-th derivative of a function up to an error of order $\cO(\epsilon^{2(k+1)})$ by evaluating the function at $2\lfloor \frac{a+1}{2}\rfloor+2k$ points. This immediately yields:
\begin{lemma}\label{lem:betterdeg}
For $a,L\in \N$ and $k\in\N_0$ the weight states $\ket{\chi_{a,L}}$ admit a representation of symmetric border rank  $2\lfloor \frac{a+1}{2}\rfloor+2k$, $\ket{\chi_{a,L,k}(\epsilon)} $,  s.t.
\begin{align*}
\|\epsilon^{-a}\ket{\chi_{a,L,k}(\epsilon)} -\ket{\chi_{a,L}}\|=\cO(\epsilon^{2(1+k)}).
\end{align*}
\end{lemma}
\begin{proof}
This immediately follows from the central finite difference formulas together with the representation of the weight states given in~\eqref{equ:representationderivative}.
\end{proof}
Moreover, it is possible to explicitly compute central difference formulas for a desired order of the derivative and accuracy by solving a system of linear equations~\cite{Quarteroni_2006}.
One corollary of Lemma~\ref{lem:betterdeg} is that, even though the weight states do not admit exact system-size independent TI representations as matrix product states, the accuracy with which we can approximate them increases exponentially with the bond dimension.

In principle these representations can be used to obtain more efficient approximate contractions schemes for states in bTNS. Indeed, by tensoring with a state with $\epsilon^{-a}\ket{\chi_{a,L,k}(\epsilon)}$ for $k$ large enough instead of $\ket{\chi_{a,L}}$ for small enough $\epsilon$ will lead to approximate contraction schemes that do not depend on $L$ and do not blow up the bond dimension of the underlying states. But it remains unclear at this stage how numerically stable these methods are.

\section{Stable Interpolation}\label{sec:stableinterpolation}
In this section, we discuss the stability result related to the border rank contraction strategy presented in the main text to compute the expectation values of an observable for a state in the new ansatz class. Given a state $\ket{\phi} \in \bTNS^{G}_{D,a,\frakd,d}$, the border rank contraction strategy considers the state $\ket{\phi(\eps)} = \ket{\Omega^{(G)}_D} \otimes \ket{\chi_{a,\frakd,L}(\eps)}$ where $\ket{\chi_{a,\frakd,L}(\eps)}$ is a border rank expression for $\ket{\chi_{a,\frakd,L}}$ (e.g., the one arising from Proposition \ref{thm:degwweight}. Given an observable $O : (\bbC^d)^{\otimes L} \to (\bbC^d)^{\otimes L}$, define 
\begin{align}\label{equ:expectationvalueeps}
p(\eps) = \eps^{-2a}\scalar{\phi(\bar{\eps})}{O\phi(\eps)};
\end{align}
then $p(0) = \scalar{\phi}{O\phi}$ is the desired expectation value. 

Rather than evaluating $p(0)$, in the border rank strategy we recover the value of $p(0)$ via Lagrange interpolation, by evaluating $p(\eps)$ at $2(L\frakd-a) +1$ points. Expanding the polynomial, we observed that the evaluation of $p(\eps)$ can be performed in terms of $(a+1)^2$ overlaps of states in $\TNS^{G}_{D,d}$.

In the main text, we only considered exact contraction schemes to compute such overlaps. However, beyond the case of 1D systems, exact contraction of tensor network states is known to be hard~\cite{PhysRevLett.98.140506,Haferkamp_2020} and, in practice, approximate contraction schemes are used~\cite{Orus_practical}. Thus, it is of paramount importance for the applicability of the border rank contraction scheme that the interpolation step is robust: in other words, even if the values of $p(\eps)$ at the $2(L\frakd -a)$ interpolation points are only exact up to an error $\delta$, one should still be able to recover the value at $0$ up to an error $\cO(\delta)$. 

Choosing the interpolation points to be roots of unity achieves this result. 

\begin{lemma}[Stable polynomial interpolation]
Let $p:\C\to\C$ be a polynomial of degree $k$ and $z_j=e^{\frac{2\pi ij}{k}}$ for $0\leq j \leq k-1$ be $k$-th roots of unity. Let $c_0 \vvirg c_{k-1} \in\C$ be $k$ points such that there exists $\delta > 0$ with  
\begin{align}\label{equ:localapprox}
    \left|p(z_j)-c_j\right|\leq \delta.
\end{align}
Then:
\begin{align*}
    \left|p(0) -\frac{1}{k}\textsum_0^{k-1} c_j\right|\leq\delta.
\end{align*}
\end{lemma}
\begin{proof}
Via Lagrange interpolation 
\begin{align*}
    p(0)=\frac{1}{k}\textsum_0^{k-1} p(z_j).
\end{align*}
Therefore 
\begin{align*}
 \left|p(0) - \frac{1}{k}\textsum_0^{k-1} c_j\right| = \frac{1}{k}\left|\textsum_0^{k-1} ( p(z_j) - c_j ) \right| \leq \frac{1}{k}\textsum_0^{k-1}\bigl|p(z_j) - c_j\bigr| \leq \delta
\end{align*}
by a triangle inequality.
\end{proof}

This result shows that, choosing the interpolation points to be the $2aL$-th roots of unity, if the approximate contraction scheme only provides the values of $z_j^{-2a}p(z_j)$ up to an error $\delta$.
Note that $z_j^{-2a}$ is a complex number of modulus $1$, so  $c_j$ satisfying $|p(z_j)-c_j|\leq\delta$ also satisfies $|z_j^{-2a}p(z_j)-c_jz_j^{-2a}|$.
The resulting approximate value of $p(0)$ is up to an error $\delta$ as well.

\bibliographystyle{plain}
\bibliography{ansatzclass.bib}